\documentclass[9pt,twocolumn,twoside,lineno]{pnas-new}

\usepackage[super, numbers, sort&compress]{natbib}
\usepackage{hyperref}
\usepackage[T1]{fontenc}
\usepackage{tgtermes}

\usepackage{float}
\usepackage{xcolor}
\usepackage{subcaption}
\usepackage{dcolumn}
\usepackage{bm}
\usepackage{multirow}
\usepackage{array}
\usepackage{siunitx} 
\usepackage{booktabs}
\usepackage{calligra}
\DeclareMathAlphabet{\mathcalligra}{T1}{calligra}{m}{n}
\DeclareFontShape{T1}{calligra}{m}{n}{<->s*[2.2]callig15}{}
\newcommand{\ra}[1]{\renewcommand{\arraystretch}{#1}}

\newcolumntype{P}[1]{>{\centering\arraybackslash}p{#1}}

\AtBeginDocument{
\heavyrulewidth=.08em
\lightrulewidth=.05em
\cmidrulewidth=.03em
\belowrulesep=.65ex
\belowbottomsep=0pt
\aboverulesep=.4ex
\abovetopsep=0pt
\cmidrulesep=\doublerulesep
\cmidrulekern=.5em
\defaultaddspace=.5em
}

\templatetype{pnasresearcharticle} 

\title{Entropy Production on Cooperative Opinion Dynamics}

\author[a, b]{Igor V. G. Oliveira}
\author[c]{Chao Wang}
\author[d]{Gaogao Dong}
\author[e]{Ruijin Du}
\author[f]{Carlos E. Fiore}
\author[g]{H. Eugene Stanley}
\author[a, b, g]{Andr\'e L. M. Vilela}

\affil[a]{F\'isica de Materiais, Universidade de Pernambuco, Recife, PE 50720-001, Brazil}
\affil[b]{Departamento de F\'isica, Universidade Federal de Pernambuco, Recife, PE 50670-901, Brazil}
\affil[c]{College of Economics and Management, Beijing University of Technology, Beijing, 100124, China}
\affil[d]{School of Mathematical Sciences, Jiangsu University, Zhenjiang, 212013, China}
\affil[e]{Center of Energy Development and Environmental Protection, Jiangsu University, Zhenjiang, 212013, China}
\affil[f]{Instituto de F\'isica, Universidade de S\~ao Paulo, S\~ao Paulo, SP, 05314-970, Brazil}
\affil[g]{Center for Polymer Studies and Department of Physics, Boston University, Boston, MA 02215, USA}

\leadauthor{Igor V. G. Oliveira} 

\onecolumn
\begin{abstract}
As one of the most widespread social dynamics, cooperative behavior is among the most fascinating collective phenomena. Several animal species, from social insects to human beings, feature social groups altruistically working for a common benefit. This collaborative conduct pervades the actions and opinions of individuals, yielding strategic decision-making between political, religious, ethnic, and economic social puzzles. Here, we explore how cooperative behavior phenomena impact collective opinion dynamics and entropy generation in social groups. We select a random fraction $f$ of community members as collaborative individuals and model the opinion dynamics using a social temperature parameter $q$ that functions as a social anxiety noise. With probability $q$, regular individuals oppose their companions about a social decision, assuming group dissent. Collaborative agents experience a reduced effective social noise $\mu q$, where $0 < \mu < 1$ is the social anxiety noise sensibility parameter that enhances social validation. We perform numerical simulations and mean-field analysis and find the system undergoes nonequilibrium order-disorder phase transitions with expressive social entropy production. Our results also highlight the effects of an individual social anxiety attenuation level in enhancing group consensus and inducing exuberant collective phenomena in complex systems.
\end{abstract}

\dates{\today}

\begin{document}

\maketitle
\thispagestyle{firststyle}
\ifthenelse{\boolean{shortarticle}}{\ifthenelse{\boolean{singlecolumn}}{\abscontentformatted}{\abscontent}}{}


\twocolumn
\dropcap{I}n light of the pervasive influence of technology, the diverse and significant challenges surrounding information dissemination have propelled intense scientific research into Sociophysics models. Several dynamics regarding opinion formation on regular and complex networks were widely proposed to investigate social, financial, and professional interactions in groups of individuals or societies. Such physical models can capture the main features of complex collective phenomena in real societies. Similar to condensed matter systems, different opinion models exhibit intense critical dynamics and nontrivial nonequilibrium phase transitions \cite{ball2002physical, schweitzer2018sociophysics, galam2008sociophysics, galam1991towards, yeomans1992statistical, galam1997rational, tome2009role, schawe2022, petri2019, centola2010, stauffer2006biology, horsevad2022, galam1999application, sznajd2000opinion, dreu2019, nyczka2013anticonformity, sznajd2014person, stone2011critical, sznajd2005left, sznajd2003effective, galam2000individual, de1992isotropic, de1993isotropic}.

Within the Sociophysics framework, the majority-vote model is an agent-based representation of interacting individuals in a contact network \cite{de1992isotropic, de1993isotropic, chen2015critical, chen2020non,  lima2022diffusive, campos2003small, choi2019majority, krawiecki2018majority, pereira2005majority, lima2007majority, crokidakis2012impact, vilela2009majority, vilela2012majority, vilela2017small, lima2013majority, vieira2016phase, encinas2019, wu2010majority, vilela2021majority, vilela2018effect, santos1995anisotropic, costa2005continuous}. The model consists of a system of agents that hold opinions for or against some issue, and the stochastic variable $\sigma_i$, which assumes one of the two values $\pm 1$, represents the opinion of an individual $i$ at a given time. The majority-vote model evolves by an inflow dynamics, where each agent agrees with the majority of its neighbors with probability $1 - q$ and disagrees with chance $q$. The quantity $q$ is called the noise parameter of the model, and it relates to a level of social anxiety, or social temperature, of the system.

Among several variations of this model, we highlight the investigation of majority-vote dynamics under the framework of random graphs and complex networks of interactions. In these studies, the authors find that group ordering, or opinion polarization in a society, is strongly related to the number of interacting neighbors \cite{choi2019majority, krawiecki2018majority, campos2003small, pereira2005majority, lima2007majority, crokidakis2012impact}, while additional investigations focus on social dynamics of systems composed of heterogeneous agents \cite{vilela2009majority, vilela2012majority, vilela2017small, lima2013majority, vieira2016phase}.

Inspired by real-world social group behavior, scientists developed further generalizations of the model, such as the three-state interpretation and different opinion functions, under the influence of regular and complex networks \cite{vilela2018effect, santos1995anisotropic, costa2005continuous, melo2010phase, lima2012three, Vilela2020three}. Nonetheless, based on opinion dynamics, examinations of this model rendered insights on second-order phase transitions, proposing criteria for the volumetric scaling of physical quantities at the critical point, yielding a universal relation for critical exponents regardless of the structure of the interaction network \cite{Vilela2020three}. Recent studies on the economic behavior of brokers in financial markets reproduced real-world market features apprised by majority-vote dynamics \cite{vilela2019majority, zubillaga2022three, granha2022opinion, zha2020opinion}.

Cooperative behavior is one of the most widespread collective social phenomena that still challenge scientists. Several animal species, from insects to human beings, exhibit social groups working for a joint benefit. Typical cooperative behavior, such as group hunting and reciprocity protection, makes species more competitive. Without this phenomenon, social institutions, non-governmental organizations, governments, culture, education, transport, health systems, among others, could be unattainable. Collaborative manners permeate the actions and opinions of individuals, imbuing strategic decision-making related to social dilemmas such as political, religious, ethnic, and economic challenges \cite{capraro2013model,de2014cooperative,pennisi2005did}.

In this paper, we design an anisotropic social model to investigate the influence of cooperative voters on group opinion evolution. We propose two types of individuals, collaborative and regular, who exhibit different chances to adopt the dominant opinion expressed in a social group. We introduce a parameter $\mu$ $\in (0, 1)$, named noise sensibility, to the standard majority-vote model to yield a distinct influence of social anxiety over individuals. Hence, a cooperative individual is under an effective attenuated social temperature $\mu q$, while a regular individual is subject to the regular noise $q$. 

Our results show that the consensus is strongly related to the number of collaborative individuals and noise sensibility. Numerical and analytical results add a significant new twist to the remarkable observation of the entropy flux of the mean-field majority-vote model \cite{tome2012entropy, hawthorne2023, harunari2019}. We achieve a general expression for isotropic and anisotropic cases and verify our results using numerical simulations in the mean-field formulation. 



\section*{Results} 
\subsection*{Model} \label{noiseaniso} In the isotropic majority-vote model (MVM), each agent occupies a node $i$ of a given network of social interactions. A spin variable $\sigma_{i}$ represents the opinion of the agent $i$ about a particular subject or in a referendum in an instant $t$. In the isotropic version, an individual is under a probability $1-q$ that its opinion $\sigma_{i}$ follows the majority state of its interacting neighbors while assuming the minority state with probability $q$ \cite{de1992isotropic, de1993isotropic}.

In this work, we analyze a square lattice opinion network with $L^2$ nodes, where a randomly chosen fraction $f$ of agents have noise sensibility $0 < \mu < 1$, addressing the behavior of the cooperative individuals. In contrast, the complementary fraction $1 - f$ of regular voters follow the standard majority-vote dynamics, i.e., $\mu = 1$. Thus, for noise level $q$, we denote the flipping probability of a given opinion $\sigma_{i}$ as 

\begin{equation}
w_{i}(\sigma) = \frac{1}{2}
     \left [ 1- (1-2\mu_i q) \sigma _{i} S \left ( \sum_{\delta = 1}^4 \sigma_{i+\delta  } \right ) \right ], 
\label{w} 
\end{equation} 
\\
the summation runs over all the four first neighboring opinions that influence the individual $i$ and $S(x)$ stands for the signal function, where $S(x) = -1, 0, 1$ for $x < 0$, $x = 0$, and $x > 0$, respectively. Furthermore,

\begin{equation}
\mu_{i} = \begin{cases}
 \displaystyle \mu, & \text{ if } i \textrm{ is a cooperative agent}.  \\
 \displaystyle 1, & \text{ if } i \textrm{ is a regular agent}.
\end{cases} \label{mu} 
\end{equation} 
\\
That is, a cooperative individual agrees with the majority with probability $1 - \mu q$, and disagrees with probability $\mu q$. Thus, noise sensibility $\mu < 1$ increases the agreement probability by attenuating the effect of the noise parameter $q$ on the society.

The cooperative majority-vote dynamics with $f = 0$ capture the behavior of the isotropic majority-vote model with noise \cite{de1992isotropic, de1993isotropic}. For $f = 1$, all individuals are cooperative, and the system also behaves as the standard MVM under the linear transformation $q \longrightarrow q/\mu$. In contrast, highlighting the effects of the noise sensibility $\mu$, we recover the standard flip probability of the isotropic MVM when $\mu = 1$, in which all the agents are under the influence of the same social temperature $q$. The case for $\mu = 0$ corresponds to a bimodal distribution of noise, where a fraction $f$ of the individuals are noiseless, always agreeing with its nearest interacting neighbors, scrutinized in previous investigations \cite{vilela2012majority, vilela2017small}. In this research, we perform numerical Monte Carlo simulations and a mean-field analytical procedure for the general cases of $0 < f < 1$ and $0 < \mu < 1$.

\begin{figure*}[ht]
\begin{center}
\includegraphics[width=0.8\linewidth]{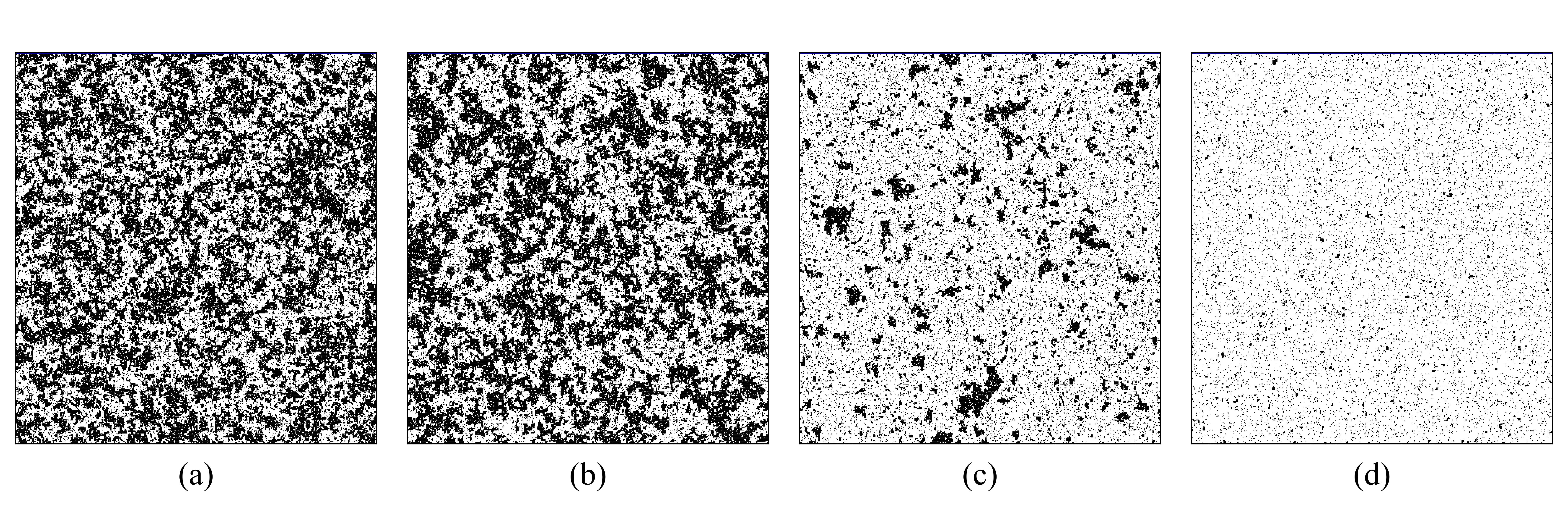}
\caption{\textbf{Snapshots of simulation on a square network.} Examples of the trajectory of a single realization with $L = 500$, $q = 0.1$ and noise sensibility $\mu = 0.5$. \textbf{a} cooperative fraction $f=0.00$, \textbf{b} $f=0.20$, \textbf{c} $f=0.50$, and \textbf{d} $f=1.00$. Increasing $f$ promotes social system consensus. White (black) dots represent $+1$ ($-1$) opinions.
}
\label{snapshots}
\end{center}
\end{figure*}
\section*{Cooperative stationary dynamics} \label{resdis}
\subsection*{Numerical results} 
To investigate the critical behavior of the model, we consider the order parameter $m$ given by

\begin{equation}\label{averageopinion}
    m = \frac{1}{L^{2}}
    \left|\sum_{i=1}^{L^{2}} \sigma_i\right|.
\end{equation}

\noindent We also consider magnetization $M_L(q,\mu,f)$, magnetic susceptibility $\chi_L(q,\mu,f)$, and Binder fourth-order cumulant $U_L(q,\mu,f)$

\begin{equation}
    M(q,\mu,f, L) = \left\langle\left\langle
    m\right\rangle_{t}\right\rangle_{c}, 
     \label{M}
\end{equation}

\begin{equation}
    \chi_L(q,\mu,f)= L^{2} \left[ \langle\langle m^2\rangle_{t}\rangle_{c}-{\langle\langle m\rangle_{t}\rangle_{c}^2} \right], 
    \label{X}
\end{equation}

\begin{equation}
    U_L(q,\mu,f)= 1-\frac{\langle\langle m^4\rangle_{t}\rangle_{c}}{3{\langle\langle m^2\rangle_{t}\rangle_{c}^2}},
    \label{U}
\end{equation}
where $\left \langle... \right \rangle_{t}$ represents time averages taken in the stationary regime, and $\left \langle... \right \rangle_{c}$ stands for configurational averages taken over independent realizations. 

We perform Monte Carlo simulations on square lattice networks with $L$ ranging from $40$ to $200$ and periodic boundary conditions. One Monte Carlo step (MCS) corresponds to the trial of updating $N$ opinions randomly chosen accordingly to (\ref{w}). Next, we discard $2 \times 10^4$ MCS to allow the system to reach the steady state and take the time averages over the subsequent $10^5$ MCS. We repeat the process up to $100$ independent samples to compute configurational averages. In our results the statistical uncertainty is smaller than the symbol size.

In Figure \ref{snapshots}, we deliver snapshots of simulations for the system with $L = 500$, $q = 0.1$ and $\mu = 0.5$, for different values of collaborative fraction $f$: (a) $0.00$, (b) $0.20$, (c) $0.50$ and (d) $1.00$. In this visual simulation, white and black dots represent opinions $+1$ and $-1$, respectively. The white area, containing a giant cluster of individuals with opinion $+1$, rises with $f$, illustrating the effects of $\mu < 1$ in promoting social order.

\begin{figure*}[ht]
\begin{center}
 \includegraphics[width=1.0\linewidth]{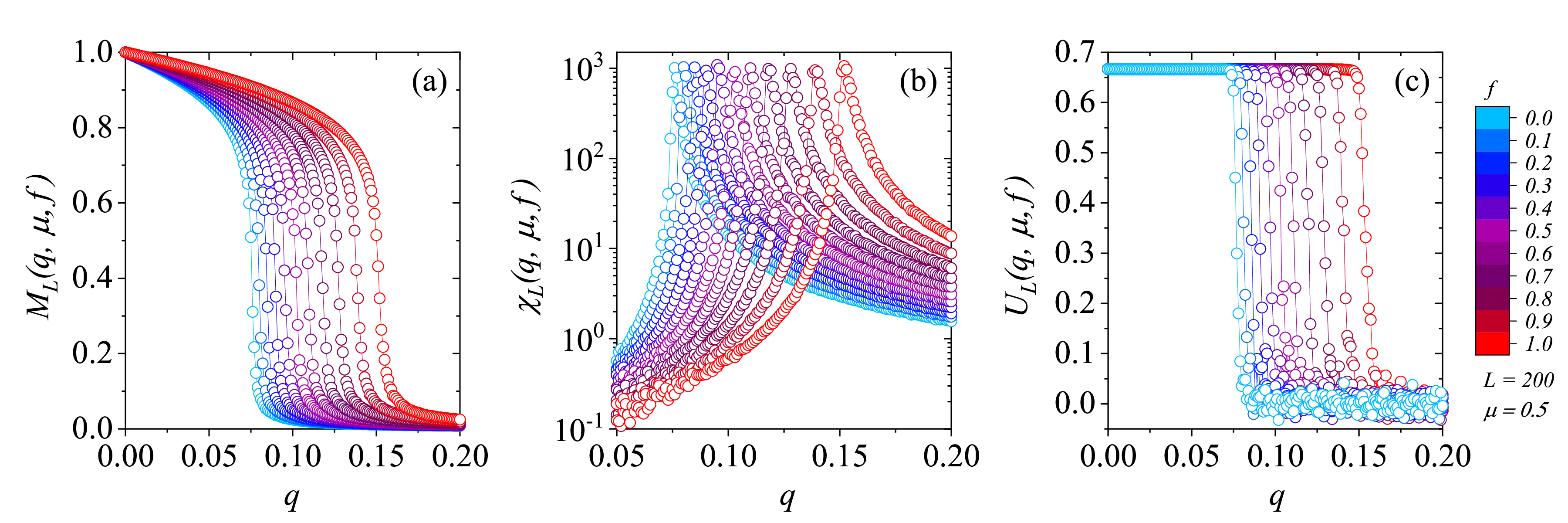}
\caption{\textbf{Stationary averages of the cooperative majority-vote opinion dynamics.} Square lattice simulations for $L = 200$, $\mu = 0.5$, and several values of $f$. Noise dependence of \textbf{a} average opinion \textbf{b} susceptibility, and \textbf{c} Binder cumulant. From left to right, $f$ increases from $0.0$ to $1.0$ with $\Delta f = 0.1$ increments. The lines are guides to the eyes.}
	\label{MXU}
\end{center}
\end{figure*}

\begin{figure*}[htpb]
\begin{center}
 \includegraphics[width=1.0\linewidth]{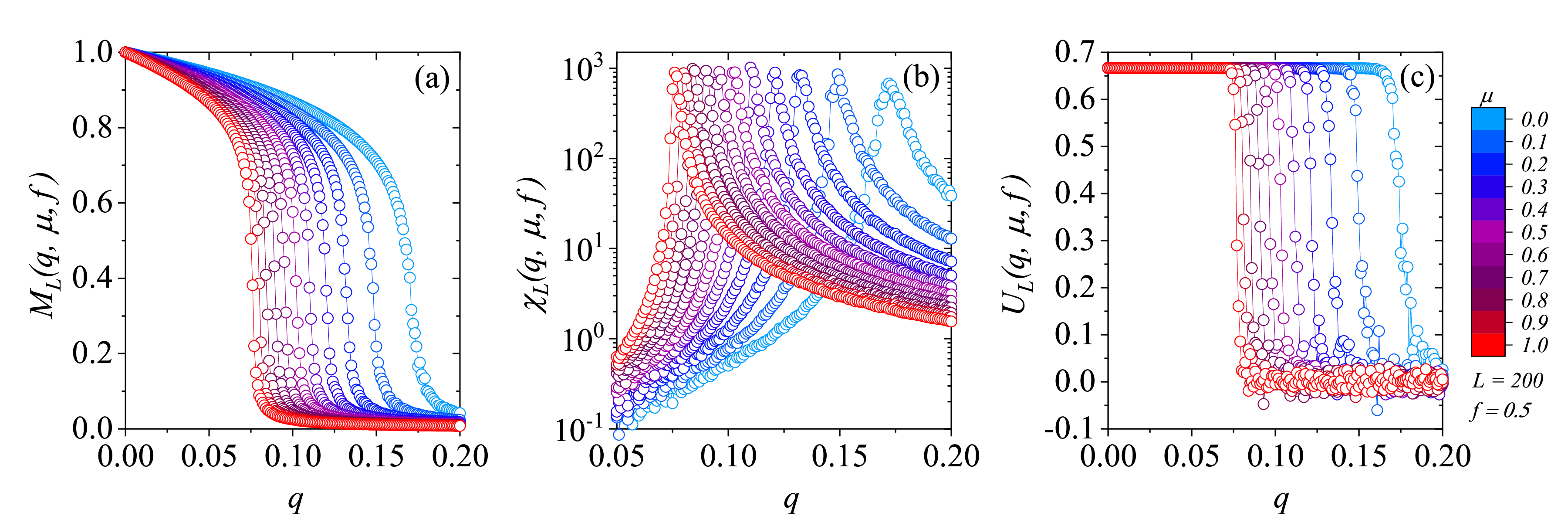}
\caption{\textbf{Effects of the intensity of cooperative behavior on consensus robustness.} \textbf{a} magnetization $M(q, c, f)$, \textbf{b} magnetic susceptibility $\chi(q, c, f)$ and \textbf{c} Binder fourth-order cumulant $U(q, c, f)$ for diverse values of $\mu$. From right to left, $\mu$ changes from $0.0$ to $1.0$ with $\Delta \mu = 0.1$ and $f = 0.5$. The lines are guides to the eyes.}
	\label{MXUMU}
\end{center}
\end{figure*}

Figure \ref{MXU} illustrates the effects of different fractions of cooperative individuals $f$ with noise sensibility $\mu = 0.5$. In this case, the cooperative agents have a $50\%$ boosted chance of agreeing with most neighbors for a given value of $q$. We plot (a) magnetization $M_L(q,\mu,f)$, (b) susceptibility $\chi_L(q,\mu,f)$, and (c) Binder cumulant $U_L(q,\mu,f)$ versus social temperature $q$ for $L=200$ and several values of $f$. For small $q$, $M(q,\mu,f) = O(1)$ indicates the ordered phase of the social system, with dominance of one opinion. By increasing social temperature $q$, $M_L(q,\mu,f)$ continuously decreases to zero for all values of the cooperative fraction $f$.

In the region where $M_L(q,\mu,f) \sim 0$, the system exhibits two opinions approximately in the same share, and the community does not support consensus. The system undergoes a second-order phase transition near a critical temperature $q_c(\mu, f)$, where the magnetic susceptibility $\chi_L(q,\mu,f)$ exhibits a maximum and the Binder cumulant $U_L(q,\mu,f)$ decreases swiftly. We remark that the critical noise value is an increasing function of the cooperative fraction $f$ since such agents improve consensus.

In Fig. \ref{MXUMU}, we study how different intensities of the cooperative behavior phenomena influence consensus when half of the community is collaborative $f = 0.5$. We investigate the behavior of (a) magnetization $M_L(q,\mu,f)$, (b) susceptibility $\chi_L(q,\mu,f)$, and (c) Binder cumulant $U_L(q,\mu,f)$ as a function of $q$ for $L=200$ and diverse values of the noise sensibility $\mu$. Decreasing $\mu$ stimulates the individuals to cooperate, reinforcing robustness to opinion disorder. Consequently, we observe that the critical noise $q_c(\mu, f)$ is a monotonically decreasing function of the noise sensibility $\mu$ for a non-zero fraction of cooperative agents.

\subsection*{Phase diagram} 
\begin{figure}[htp]
    \begin{center}
    \includegraphics[width=1.0\linewidth]{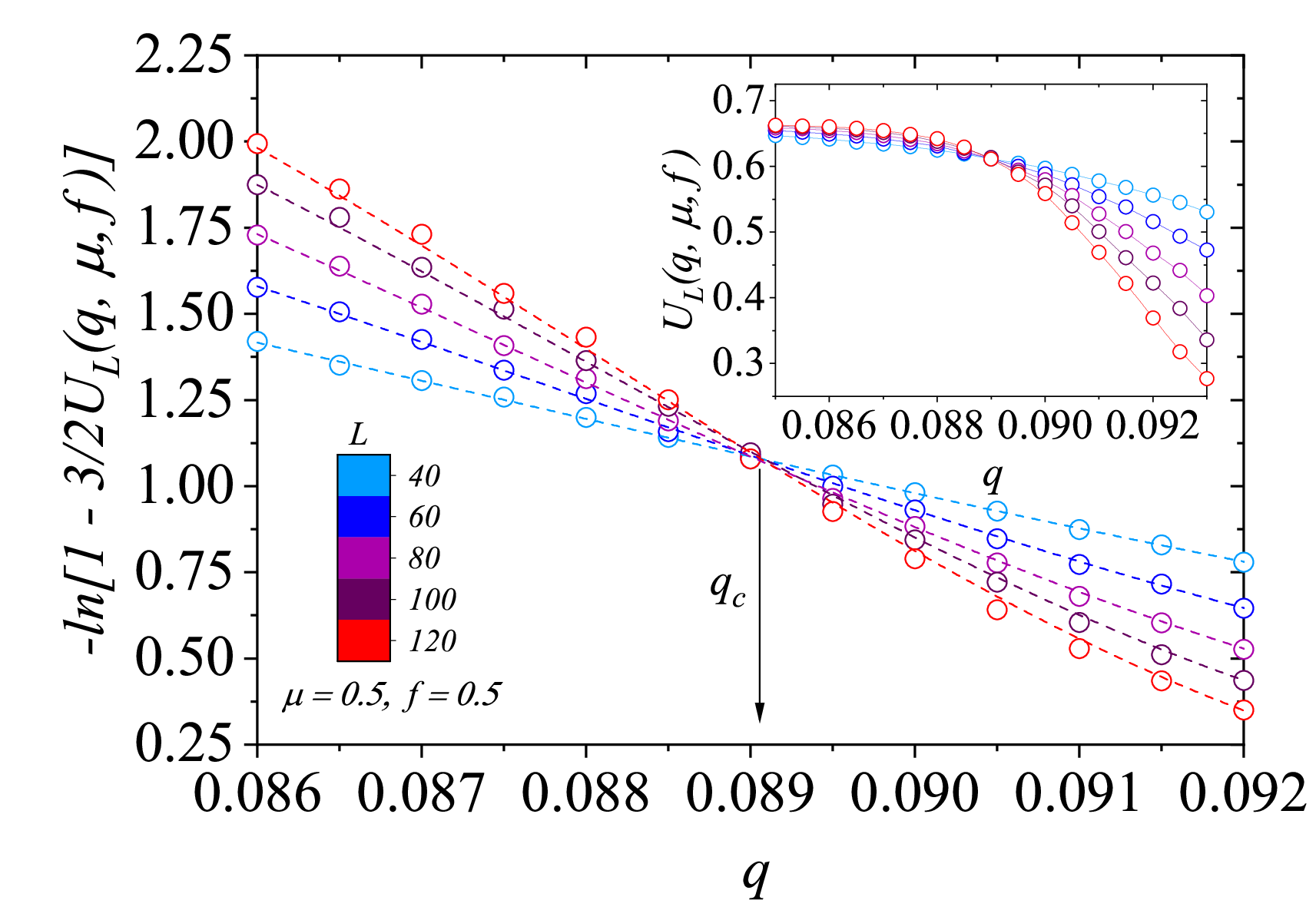}
    \caption{\textbf{Critical noise estimative.} Binder fourth-order cumulant $U_L(q,\mu,f)$ for $\mu = 0.5$ and cooperative fraction $f=0.3$. The point where the curves for different sizes $L$ intersect provides an estimate for critical social temperature $q_{c}(\mu, f)$ in the thermodynamic limit $N \rightarrow \infty$. The dashed lines are cubic fits of the data points in critical region, and the continuous lines are guides to the eyes.} 
\label{U(L)}
    \end{center}
\end{figure}

\begin{figure}[htp]
    \begin{center}
    \includegraphics[width=1.0\linewidth]{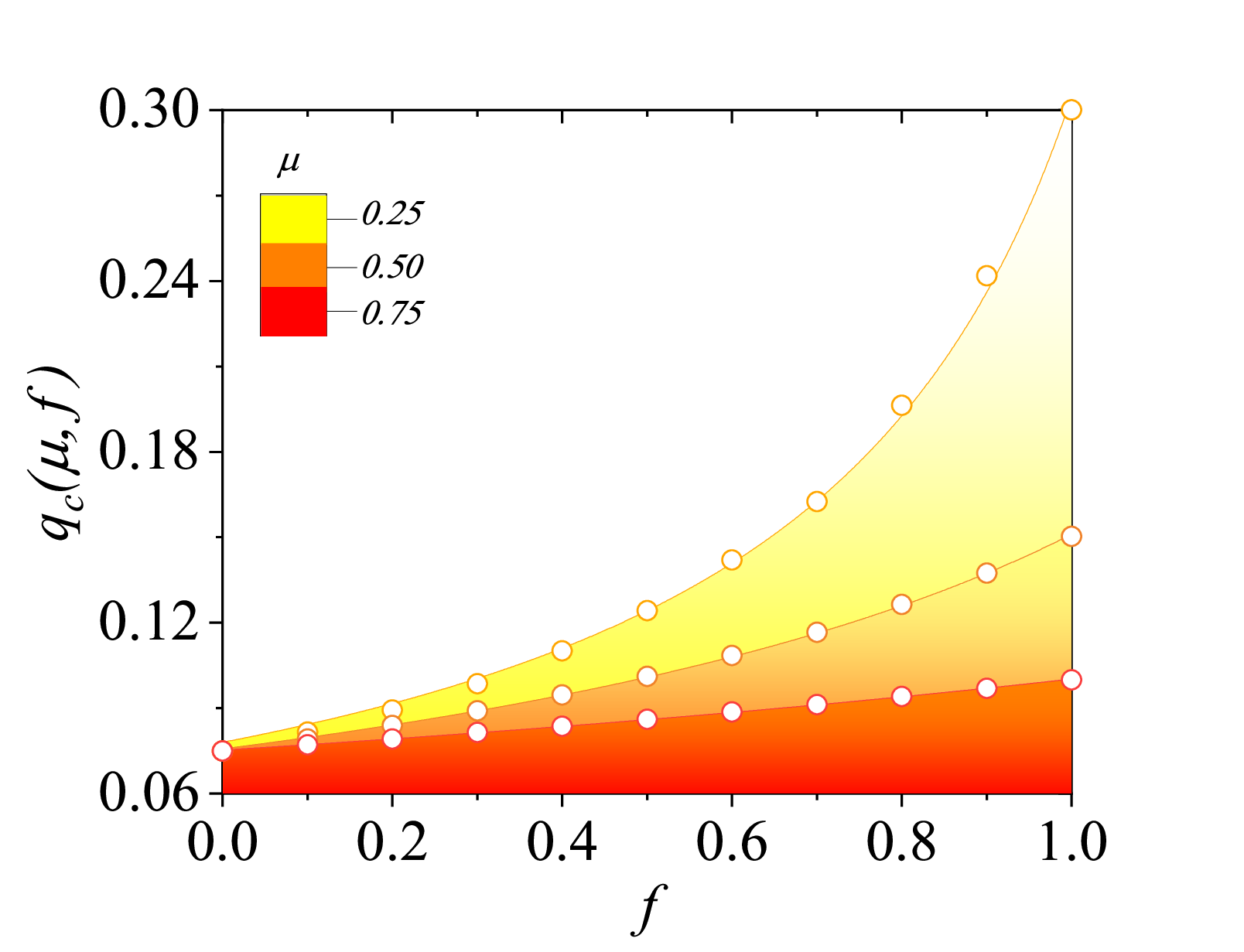}
    \caption{\textbf{Phase diagram of cooperative majority-vote opinion dynamics.} The curves are descriptions of the phase boundary that separates the ordered and disordered phases for different values of noise sensibility $\mu$. Circles represent the numerical estimates of critical points $q_{c}(\mu,f)$, obtained by the crossing point of the Binder cumulant curves for different system sizes. Lines are fits from  \eqref{qc2dequation}.}
\label{dia2d}
    \end{center}
\end{figure}

To obtain a precise estimate of the critical social temperature $q_{c}(\mu,f)$ in the thermodynamic limit $N \to \infty$, which is independent of the society scale $L$, we calculate the Binder fourth-order cumulant for each pair $(\mu, f)$ with different system sizes. In Figure \ref{U(L)}, we exemplify this method by displaying the Binder cumulant for $\mu = 0.5$ and $f = 0.3$. We estimate the critical noise value $q_{c}(\mu, f)$ from the intersection point of Binder curves for different sizes $L$, since $U$ does not depend on the system size only at $q = q_{c}(\mu, f)$. We find $q_{c}(\mu, f) = 0.0891(2)$ for $\mu = 0.5$ and $f = 0.3$. In Table \ref{square_lattice_qcs}, we summarize the results for the same process employing other values of $f$ and $\mu$, rendering the phase diagram shown in Fig. \ref{dia2d}.

\begin{table*}\centering
\ra{0.4}
\setlength{\tabcolsep}{0.35cm}
\caption{Critical social temperatures on square lattices and mean-field estimates $q_{c}(\mu,f)$ and $q_{c}^{MF}(\mu,f)$, respectively, as a function of $f$ and $\mu$ for the cooperative majority-vote dynamics.}
\begin{tabular}{c c c c c c c} 
\cellcolor{red!40} $f$  &  \cellcolor{orange!50}$q_{c}(\mu=1/4)$  & \cellcolor{orange!50}$q_{c}(\mu=1/2)$ & \cellcolor{orange!50}$q_{c}(\mu=3/4)$ &  \cellcolor{blue!20}$q_{c}^{MF}(\mu=1/4)$  & \cellcolor{blue!20}$q_{c}^{MF}(\mu=1/2)$ & \cellcolor{blue!20}$q_{c}^{MF}(\mu=3/4)$\\ 
\\0.0 & 0.0750(1) & 0.0750(3)  & 0.0750(1) & 0.1665(1)   & 0.1665(1)  & 0.1664(3)\\
\\0.1 & 0.0816(2) & 0.0791(2)  & 0.0771(1) & 0.1802(1)   & 0.1753(1)  & 0.1711(2)\\
\\0.2 & 0.0894(1) &  0.0839(1) & 0.0792(1) & 0.1957(3)   & 0.1851(1)  & 0.1750(1)\\
\\0.3 & 0.0986(1) &  0.0891(2) & 0.0814(1) & 0.2149(1)   & 0.1961(1)  & 0.1799(2)\\
\\0.4 & 0.1101(2) &  0.0947(2) & 0.0837(1) & 0.2376(4)   & 0.2077(3)  & 0.1848(1)\\
\\0.5 & 0.1243(1) &  0.1011(1) & 0.0861(2) & 0.2667(1)   & 0.2224(2)  & 0.1904(1)\\
\\0.6 & 0.1420(3) &  0.1085(2) & 0.0886(2) & 0.3031(3)   & 0.2373(1)  & 0.1955(1)\\
\\0.7 & 0.1626(3) &  0.1167(2) & 0.0912(2) & 0.3507(1)   & 0.2566(2)  & 0.2018(2)\\
\\0.8 & 0.1963(3) &  0.1264(2) & 0.0941(2) & 0.4163(4)   & 0.2768(2)  & 0.2076(1)\\
\\0.9 & 0.2418(3) &  0.1374(1) & 0.0970(1) & 0.5128(1)   & 0.3033(3)  & 0.2152(1)\\
\\1.0 & 0.3002(1) &  0.1503(2) & 0.1000(1) & 0.6664(1)   & 0.3332(2)  & 0.2221(1)\\

\bottomrule
\end{tabular}
\label{square_lattice_qcs}
\end{table*}

The interpolation of critical points $q_{c}(\mu,f)$ in  Fig. \ref{dia2d} generates a description of the phase boundary that separates the ordered and disordered phases for each value of the noise sensibility $\mu$. We note that consensus correlates with noise sensibility, and lower values of $\mu$ tend to yield higher values of $q_c$. Consensus robustness is also proportional to $f$ since it controls the fraction under the influence of an effective noise reduction. From the data, we propose the phase boundary lines to obey an equation of type

\begin{equation}
    q_{c}(\mu,f)= \frac{1}{a-bf},
    \label{qc2dequation}
\end{equation}

\noindent in which $a$ and $b$ are parameters that depend on $\mu$. By conducting a non-linear curve fitting using  \eqref{qc2dequation}, we estimate $[a , b] =  [12.8(5), 9.4(5)], [13.2(1) , 6.5(1)] , [13.2(1) , 3.3(1)] $,  for $\mu = 0.25, 0.50$ and $0.75$, respectively. From Table \ref{square_lattice_qcs}, we obtain $q_{c}(\mu, 0) = 1/a  \approx 0.075$, in agreement with the isotropic MVM \cite{de1992isotropic}, and $q_{c}(\mu, 1) = 1/(a-b) \approx 0.075/\mu$ as expected from previous analysis.

\begin{figure*}[htpb]
    \begin{center}
    \includegraphics[width=1.0\linewidth]{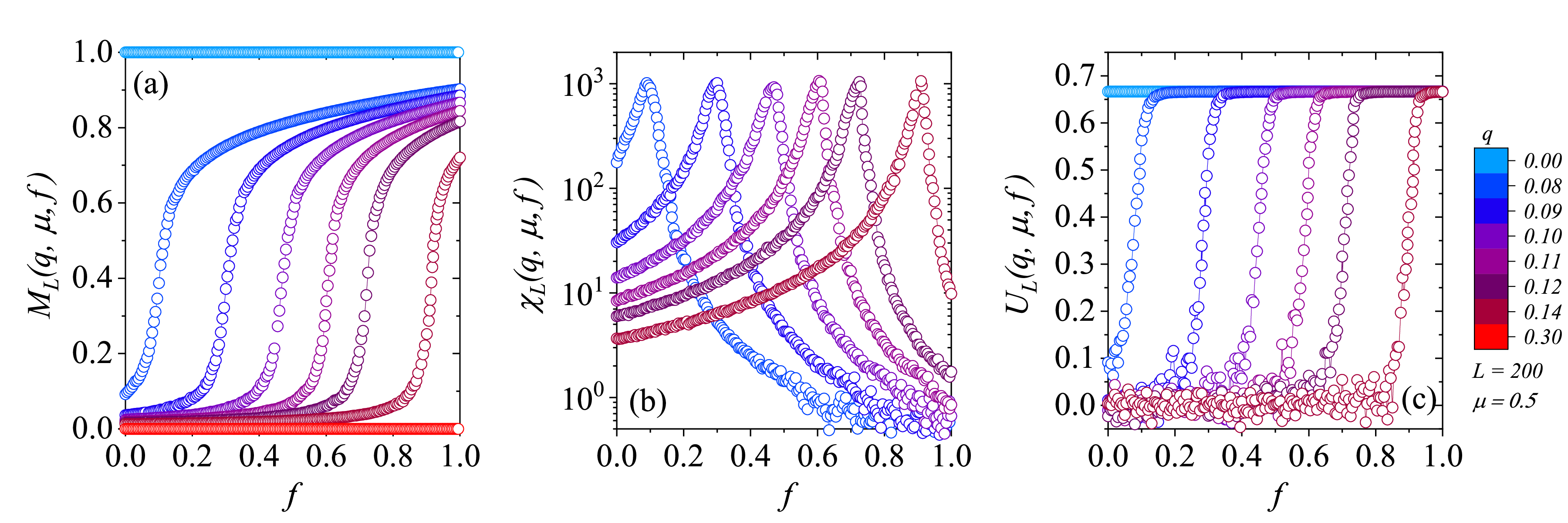}
   \caption{\textbf{Disorder-order transitions induced by cooperative agents.} In this configuration, $L = 200$ and $\mu = 0.5$ for different values of $q$. Figures \textbf{a}, \textbf{b}, and \textbf{c} stand for magnetization, susceptibility, and Binder cumulant, respectively. From left to right, $q$ = $0.08, 0.09, 0.10, 0.11, 0.12$ and $0.14$.}
    \label{slmxf}
    \end{center}
\end{figure*} 

In Figure \ref{slmxf}, we illustrate how cooperative fraction $f$  improves order for $L = 200$ and $q$ = $0.08, 0.09, 0.10, 0.11, 0.12$ and $0.14$ (from left to right). Note the system undergoes a disorder-order transition for intermediate values of $q$. High values of $f$ suppress disorder, ordering the community regardless of noise $q$. We highlight limiting cases that are independent of $f$ for $\mu = 0.5$ in Fig. \ref{slmxf}(a), where we use $q = 0$ and $q = 0.3$.

\subsection*{Critical exponents}
\begin{figure*}[htpb]
    \begin{center}
     \includegraphics[width=1.0\linewidth]{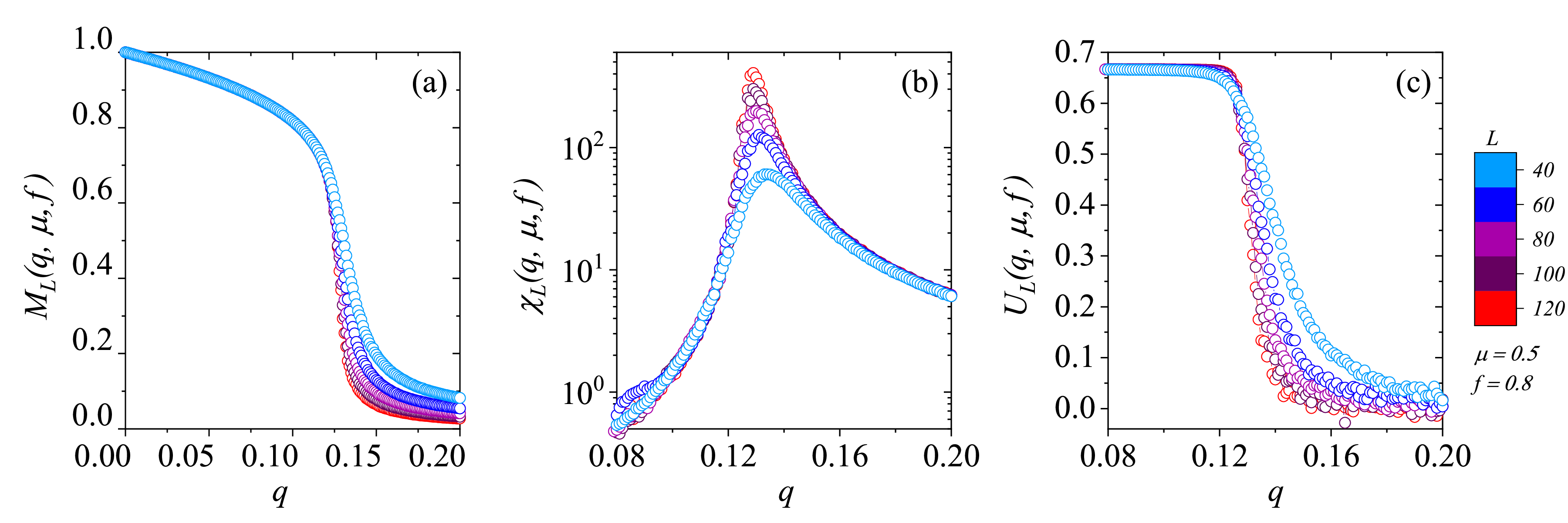}
    \caption{\textbf{Size dependence on consensus robustness versus noise parameter.} In \textbf{a} average opinion, \textbf{b} susceptibility and \textbf{c} Binder cumulant $U$ for system sizes $L = 40, 60, 80, 100$, and $120$. 
     In this result, $f=0.8$ and $\mu = 0.5$. The lines are guides to the eyes.}
\label{MXL}
    \end{center}
\end{figure*}

We examine finite-size effects on the social dynamics of the cooperative majority-vote model. In Figure \ref{MXL}, we exhibit (a) magnetization, (b) susceptibility and (c) Binder cumulant for $f=0.8$ and $\mu = 0.5$, with $L = 40, 60, 80, 100$ and $120$. Note that at the critical point $q_c(\mu, f) \approx 0.13$ (see Table \ref{square_lattice_qcs}), $M \to 0$ as $L \to \infty$, remaining non-zero for noise values below $q_c(\mu, f)$. Also, the larger $L$, the more intense the magnetization fluctuations, yielding the highest peaks observed for the magnetic susceptibilities near $q_c(\mu, f)$.

\begin{figure*}[htpb]
\begin{center}
	\includegraphics[width=1.0\linewidth]{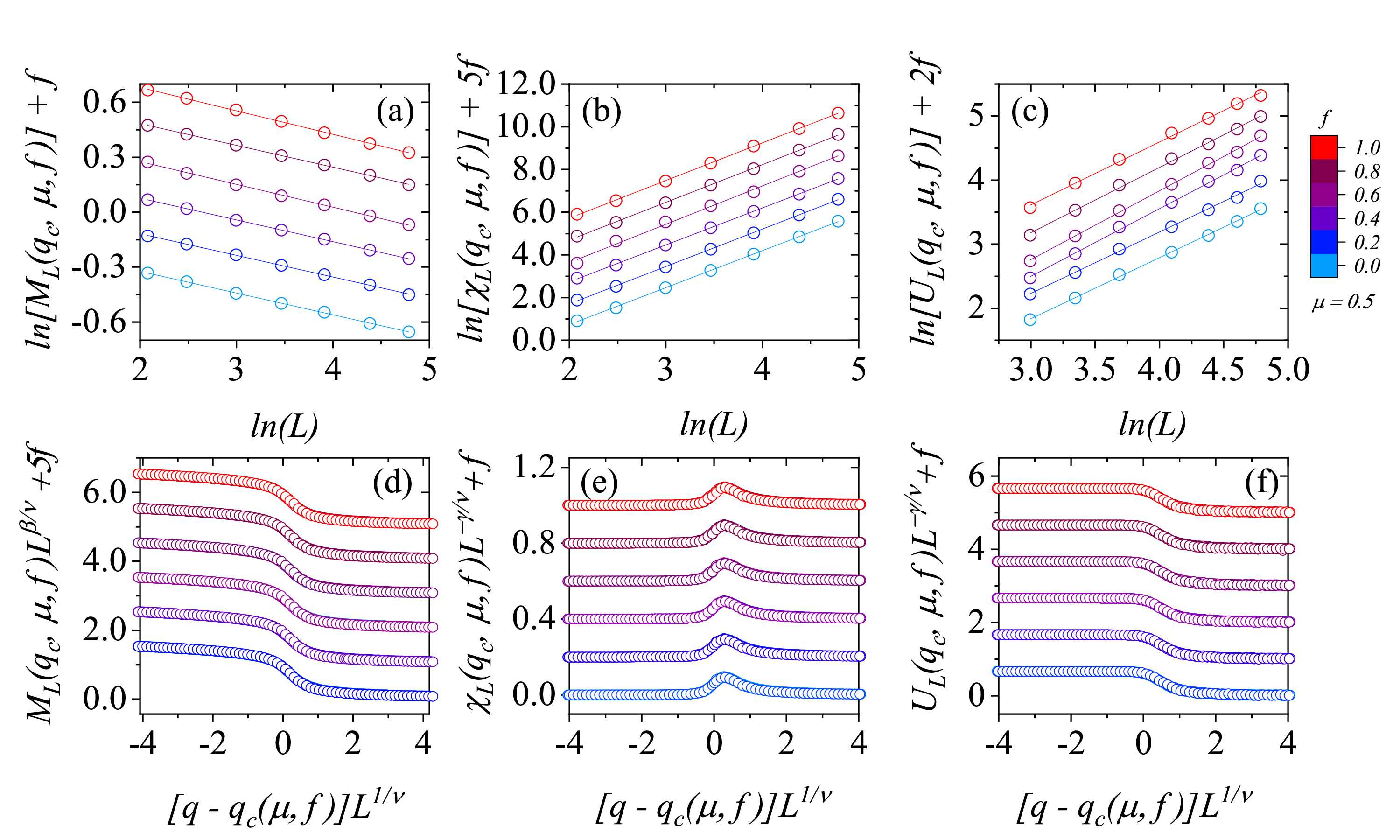} 
\caption{\textbf{Finite-size scaling analysis and universality.} \textbf{a} Magnetization, \textbf{b} magnetic susceptibility, and \textbf{c} Binder fourth-order cumulant at the critical point $q = q_{c}(\mu, f)$ as functions of linear system size $L$ in log-log scale for several values of the cooperative fraction $f$ and $\mu = 0.5$. The lines represent linear fits to the data, yielding the standard Ising model critical exponents on square lattices considering error bars. We rescale all quantities, rendering one universal curve for critical exponents $\beta /\nu = 0.125 $, $\gamma  /\nu = 1.75$, and $1/ \nu = 1$. We shift curves up to avoid overlap.} 
\label{EXPO_SQUARE}
\end{center}
\end{figure*}

\begin{figure*}[htpb]
\begin{center}
	\includegraphics[width=1.0\linewidth]{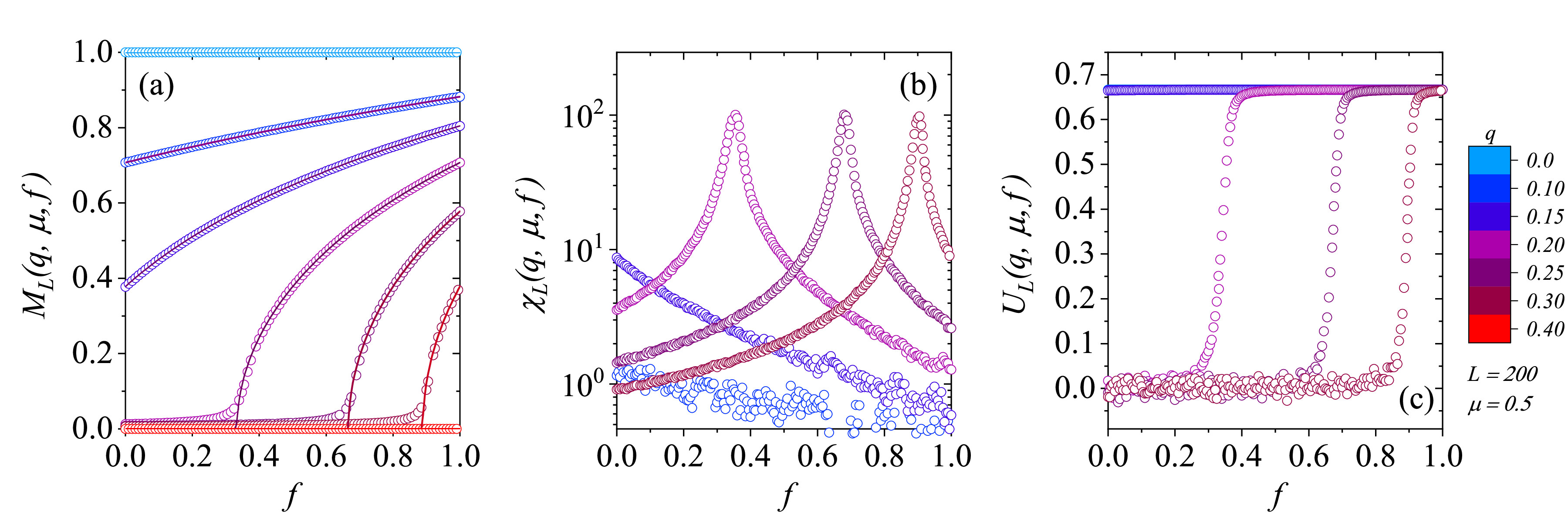} \quad
\caption{Plots of the (a) magnetization, (b) magnetic susceptibility, and (c) Binder cumulant as a function of $f$ for several values of the noise $q$ for $\mu = 0.5$. The dashed lines on (a) stand for the analytical results given by  \eqref{p8}, and the symbols are the numerical results for $20$ samples with $L= 200$.}
	\label{mxf}
\end{center}
\end{figure*}

To further analyze the behavior of $M$, $\chi$, and $U$ with system size $L$ near the critical point, we estimate the critical exponents $\beta/\nu$, $\gamma/\nu$, and $1/\nu$ that characterize the phase transition of the model. Thus, we write the following finite-size scaling relations

\begin{equation}
    M_{L}(q,\mu,f)= L^{-\frac{\beta}{\nu}}\widetilde{M} ( \varepsilon L ^{\frac{1}{\nu}}),
    \label{mr}
\end{equation}
 \begin{equation}
    \chi_{L}(q,\mu,f) = L^{\frac{\gamma}{\nu}}\widetilde{\chi}(\varepsilon L^{\frac{1}{\nu}}),
    \label{xr}
\end{equation}
 \begin{equation}
    U_{L}(q,\mu,f) = \widetilde{U}(\varepsilon L^{\frac{1}{\nu}}),
    \label{ur}
\end{equation}
where $\varepsilon = q - q_{c}(\mu,f)$ is the distance from critical noise, and the universal scaling functions $\widetilde{M}$, $\widetilde{\chi}$ and $\widetilde{U}$ depend only on scaling variable $x = \varepsilon L^{\frac{1}{\nu}}$. Accordingly, we use these equations to obtain the critical exponents of phase transitions and capture the universal behavior of magnetization, magnetic susceptibility, and Binder cumulant.

In Figure \ref{EXPO_SQUARE}, we illustrate the numerical results for (a) $M$, (b) $\chi$ and (c) $U$ versus the system size $L$ at $q_c(\mu, f)$, with $\mu = 0.5$ and several values of $f$. By measuring the linear coefficient of each line in Fig. \ref{EXPO_SQUARE}(a), Fig. \ref{EXPO_SQUARE}(b) and Fig. \ref{EXPO_SQUARE}(c), we estimate $\beta /\nu \approx 0.125 $, $\gamma  /\nu \approx 1.75$ and $1/ \nu  \approx 1$ considering the error bars. We confirm our results by performing a data collapse of rescaled versions (d) $\widetilde{M}$, (e) $\widetilde{\chi}$ and (f) $\widetilde{U}$ over the rescaled social noise using $\beta /\nu = 0.125 $, $\gamma  /\nu = 1.75$ and $1/ \nu = 1$. Despite the different behaviors observed in Figs. \ref{MXU} and \ref{MXL}, Figs. \ref{EXPO_SQUARE}(d), (e), and (f) yield a single universal curve independently on $f$.

We further investigate critical exponents for $\mu = 0.25$ and $\mu = 0.75$, and the results also supply the same set of critical exponents. We conclude that the critical exponents of the cooperative majority-vote model are the same as those in an equilibrium two-dimensional Ising model and for the isotropic majority-vote dynamics \cite{de1992isotropic}, regardless of $\mu$ and $f$. This result is under Grinstein’s criterion that states that nonequilibrium stochastic spin-like systems with up-down symmetry in regular lattices fall into the same universality class of the equilibrium Ising model \cite{grinstein1985statistical,baxter1982inversion}.

\subsection*{Mean-field analyses}
\label{meanfieldapp}
A given configuration of opinions can be denoted by $\sigma = (\sigma_{1} , \sigma_{2} , . . . , \sigma_{i} , . . . , \sigma_{N} )$, with $N = L^2$. We obtain the behavior of the stationary magnetization $m$ using the master equation that expresses the evolution of the probability $P(\sigma, t )$ of finding the system in the state $\sigma$ at a time $t$ \cite{van1992stochastic, fiore2023}

\begin{equation}
\frac{d}{dt}P(\sigma,t) =\displaystyle
     \sum_{i=1}^{N}\left [ w_{i}(\sigma ^i)P(\sigma^i,t)- w_{i}(\sigma)P(\sigma,t) \right ],
  \label{P}
\end{equation}
where the state ${\sigma^i}$ can be obtained from state ${\sigma}$ flipping the $i$-th agent's opinion, i.e., ${\sigma^i } = (\sigma_{1} , \sigma_{2} , . . . , -\sigma_{i} , . . . , \sigma_{N} )$. Factor $w_{i}$ is the flip rate of the $i$-th individual $\sigma_{i} \rightarrow  -\sigma_{i}$, given by  \eqref{w} for the cooperative voter model. From  \eqref{P}, it follows that the time evolution of the average opinion of the agent $\sigma _{i}$ is

\begin{equation}
\frac{d}{dt}\left \langle \sigma_{i} \right \rangle = -2\left \langle \sigma_{i}w_{i}  \right \rangle. 
  \label{sig}
\end{equation}
Thus, for all $Nf$ cooperative individuals, we write the following set of equations

\begin{equation}
\displaystyle\frac{d}{dt}\left \langle \sigma _{j} \right \rangle  = - \left \langle \sigma _{j} \right \rangle + \Theta_\mu \left \langle S\left ( \displaystyle
     \sum_{\delta }^{} \sigma_{j + \delta }  \right ) \right \rangle, \\ 
  \label{C1}
\end{equation}
for $j = 1, 2,..., Nf$, where we replace $w_j$ using  \eqref{w} with $\Theta_\mu = 1- 2\mu q$ and $\sigma^2_j = 1$. Similarly, for the remaining $N(1-f)$ agents, we have

\begin{equation}
\displaystyle\frac{d}{dt}\left \langle \sigma _{k} \right \rangle  = - \left \langle \sigma _{k} \right \rangle + \Theta \left \langle S\left ( \displaystyle
     \sum_{\delta }^{} \sigma_{k + \delta }  \right ) \right \rangle, \\ 
  \label{C2}
\end{equation}
where $\Theta  = 1-2q$ and $k = Nf + 1, Nf + 2, ..., N$. Adding Equations \eqref{C1} and \eqref{C2} and summing for all agents, we obtain

\begin{equation}
  \begin{aligned}
 & \sum_{i = 1}^{N} \frac{d}{dt} \left \langle \sigma_{i} \right \rangle = -\sum_{i = 1}^{N}\left \langle \sigma_{i}  \right \rangle + \\
& + N \left [ f\Theta_\mu + (1-f)\Theta \right ]\left \langle S\left ( \sum_{\delta} \sigma_{i + \delta } \right )  \right \rangle.
     \end{aligned}
   \label{p11}
\end{equation}

\noindent In the mean-field limit, a randomly chosen agent $\sigma_{i}$ interacts with four neighbors also randomly selected. Labeling these neighbors as $\sigma_{a}$,  $\sigma_{b}$,  $\sigma_{c}$ and $\sigma_{d}$, we write \cite{hawthorne2023, harunari2019}

\begin{center}
        $\displaystyle
      S\left ( \sum_{\delta} \sigma_{i + \delta } \right )  = S( \sigma_{a}+\sigma_{b} + \sigma_{c} + \sigma_{d}) $ 
\end{center}

\begin{equation}  
      \begin{aligned}
& = \frac{3}{8}( \sigma_{a}+\sigma_{b} + \sigma_{c} + \sigma_{d}) + \\
& - \frac{1}{8}( \sigma_{a}\sigma_{b}\sigma_{c}+\sigma_{a}\sigma_{b}\sigma_{d} + \sigma_{a}\sigma_{c}\sigma_{d} + \sigma_{b}\sigma_{c}\sigma_{d}).
     \end{aligned}
     \label{p2}
\end{equation}
In addition, in the stationary state, $ m \approx \left \langle \sigma_{i} \right \rangle$ and $\left \langle \sigma_{l}\sigma_{u}\sigma_{v} \right \rangle \approx   \left \langle \sigma_{l}\right \rangle \left \langle \sigma_{u}\right \rangle \left \langle \sigma_{v}\right \rangle \approx m^3$. Thus, we write 

\begin{equation}
\left \langle S\left ( \sum_{\delta} \sigma_{i + \delta } \right )  \right \rangle = \frac{m}{2}(3 - m^2).
\end{equation}
By using this result in  \eqref{p11}, we obtain 

\begin{equation}
\displaystyle\frac{d}{dt} m = m\left \{ -\epsilon -\frac{m^2}{2}\left [ f\Theta_\mu + (1-f)\Theta \right ]   \right \},
  \label{p5}
\end{equation}
where

\begin{equation}
     \epsilon = 1 - \frac{3}{2}\left [ f\Theta_\mu + (1-f)\Theta  \right ].
   \label{p6}
\end{equation} 
In the stationary state, $dm/dt = 0$. For $\epsilon > 0$, there is only one real solution, $m=0$, representing the paramagnetic state (disordered). For $\epsilon < 0$, we obtain the ferromagnetic state (ordered) solution 
\begin{equation}
m = \sqrt{\frac{2\left | \epsilon  \right |}{  f\Theta_\mu + (1-f)\Theta  }},
   \label{p7}
\end{equation} 
Then, using $\Theta_\mu = 1-2 \mu q$ and $\Theta = 1 - 2q$ and  \eqref{p6}, we write 

\begin{equation}
m = \sqrt{\frac{1 - 6q\left [ 1 - f\left (1 - \mu \right ) \right ]}{1 - 2q\left [ 1 - f\left (1 - \mu \right ) \right ]}} \equiv \sqrt{\frac{1 - 6\bar{q}}{1 - 2\bar{q}}},  
\label{p8}
\end{equation} 
with $\bar{q} = \bar{\mu}q = q[(1-f( 1-\mu)]$, valid for $q < q_c^{MF}$, the mean-field critical temperature. By imposing $m = 0$ in  \eqref{p8}, we obtain
\begin{equation}
q_{c}^{MF}(\mu,f) = \frac{1}{6\left [  1-f\left ( 1-\mu \right )\right ]}.
   \label{qc}
\end{equation} 
 Note that when $f = 0$,  \eqref{p8} yields the isotropic MVM mean-field result for $m$

\begin{equation}
m = \sqrt{\frac{1 - 6q}{1 - 2q}},
   \label{standardp8}
\end{equation} 
with $q_c^{MF} = 1/6$ \cite{tome2015stochastic}. For $f = 1$, $q_{c}^{MF} = 1 / 6\mu$ as anticipated. Additionally, near the phase transition, $m \sim ( |q - q_{c}^{MF}|)^ \beta $, and we find exponent $\beta = 1/2$, indicating that the cooperative majority-vote model should belong to the mean-field Ising universality class.

We confirm our analytical results by performing Monte Carlo simulations in the mean-field approach. In this formulation, we randomly select an agent whose four neighbors are also randomly chosen. We consider systems of $N$ agents, with $N = 1600, 3600, 6400, 10000$ and $40000$. We skip $10^3$ MCS to allow thermalization and evaluate the time averages over the next $10^5$ MCS up to $100$ different samples.

In Fig. \ref{mxf}, we show mean-field numerical estimates for (a) $M_{L}(q, \mu, f)$, (b) $\chi_{L}(q, \mu, f)$ and (c) $U_{L}(q, \mu, f)$ as functions of the fraction of collaborative individuals $f$ for several values of the noise $q$. We evaluate the magnetization numerically (circles) and compare it with the analytical solution of  \eqref{p8} (lines). The small divergences near the phase transition point result from the limited nature of the simulated network with $L=200$, whereas the analytical solution assumes the thermodynamic limit $N \longrightarrow \infty$. The maximum of each susceptibility curve in Fig. \ref{mxf}(b) denotes the critical values of $f$ that yield an order-disorder phase transition. Additionally, the critical noise $q$ necessary to vanish the order consensus increases with $f$, denoting a boost of social robustness. This behavior agrees with  \eqref{qc}, where $q_{c}^{MF}(\mu, f)$ is a monotonically increasing function of $f$.

\begin{figure}[ht!]
\begin{center}
\includegraphics[width=1.0\linewidth]{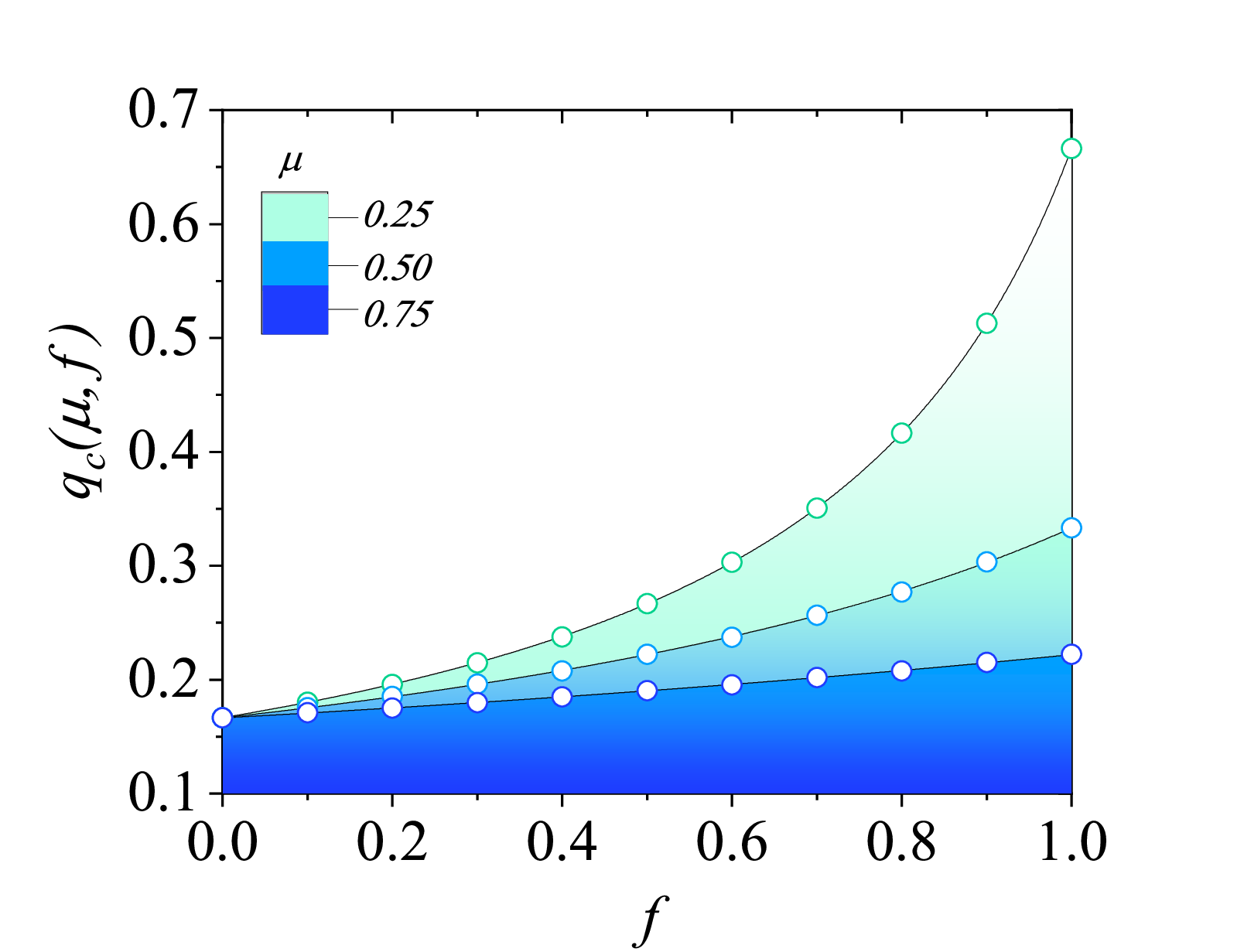} 
\caption{\textbf{Mean-field consensus-dissensus phase diagram.} Lines denote analytical solutions given by \eqref{qc}, producing the phase boundaries that separate the ordered and the disordered phases for each noise sensibility $\mu$. Circles represent numerical results for $q_{c}^{MF}(\mu, f)$, estimated by intersection points of Binder cumulant curves in mean-field simulations.}
\label{diagram_meanfield}
\end{center}
\end{figure}

\begin{figure}[ht!]
  \includegraphics[width = 1.0\linewidth]{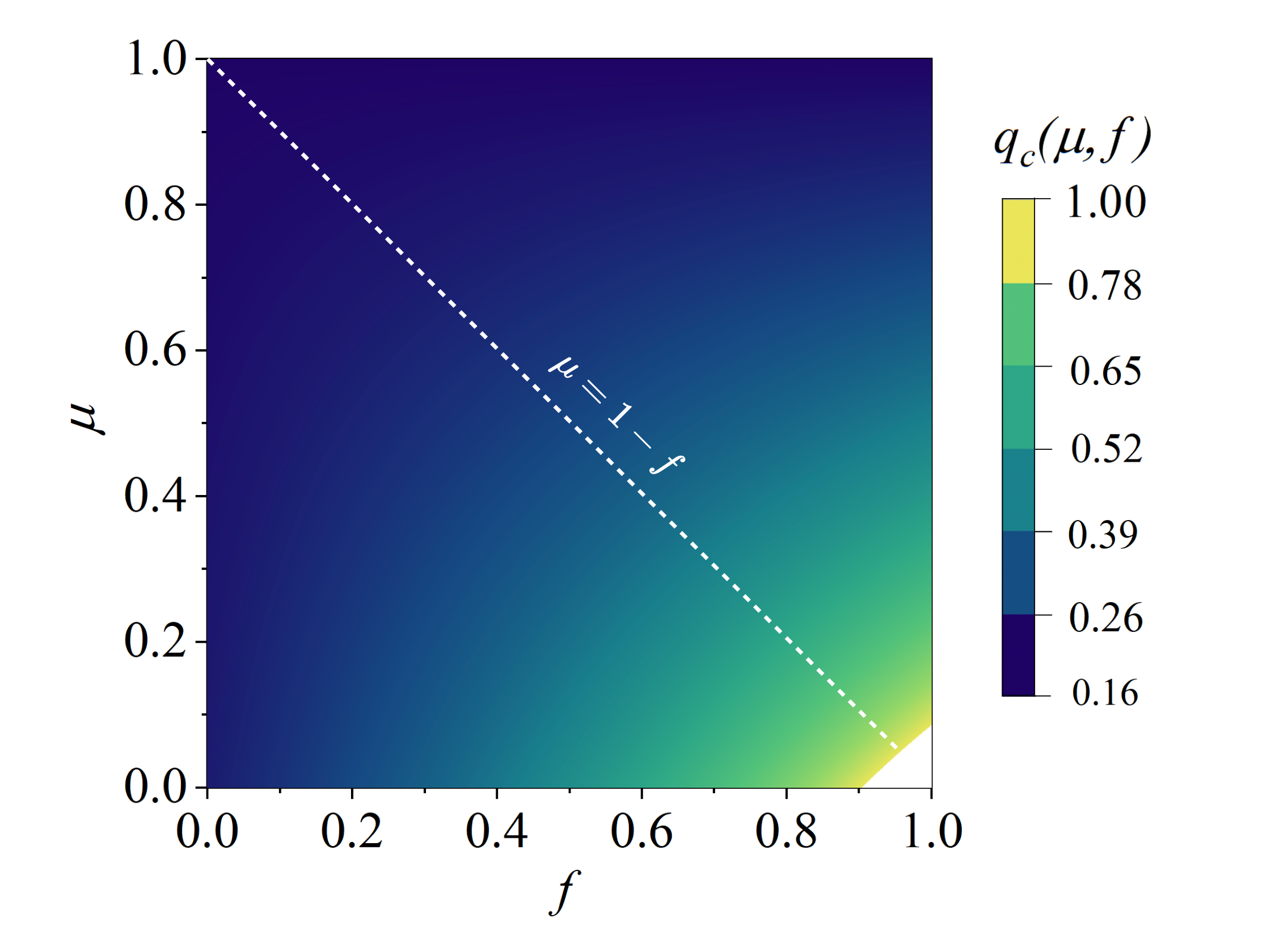}
  \caption{\textbf{Density plot for mean-field critical noise as a function of cooperative parameters.} The figure illustrates the results of \eqref{qc}. In the white region, the critical social temperature is higher than one, and consequently, the system yields a social consensus for all investigated noise values. The white dotted line corresponds to the symmetry line $\mu = 1 - f$.}
  \label{DensityQc}
\end{figure}

\begin{figure*}[ht!]
\begin{center}
 	\includegraphics[width=1.0\linewidth]{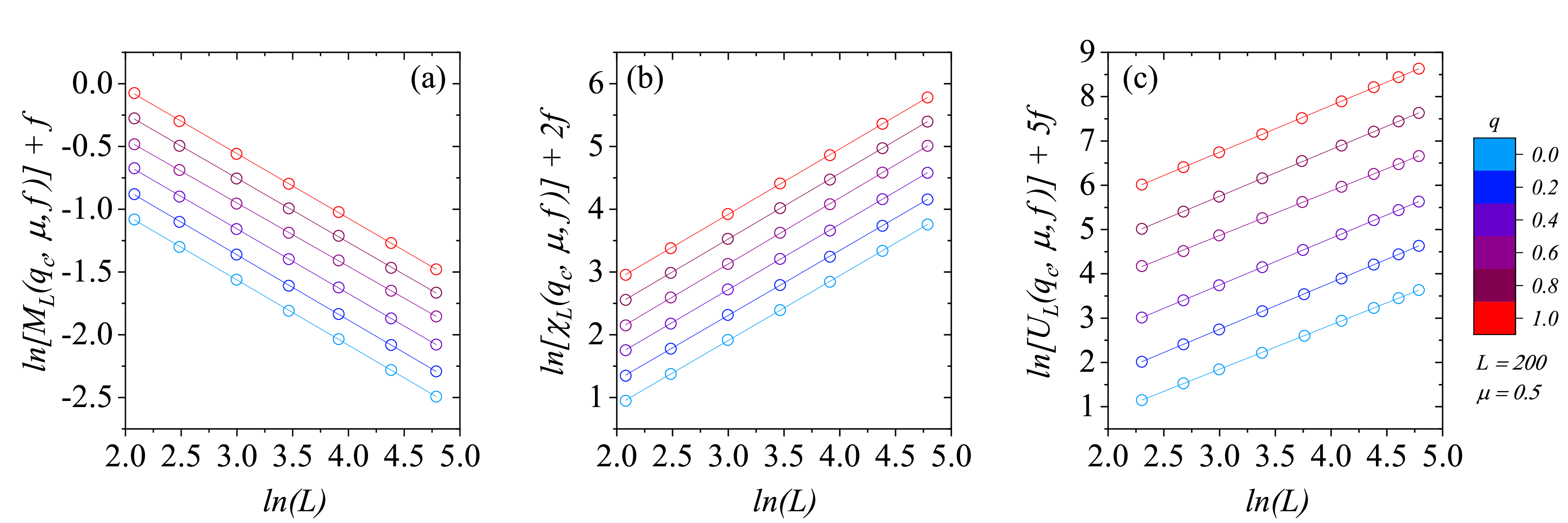} \quad
\caption{\textbf{Mean-field finite-size estimates for critical exponents.} \textbf{a} $M(q, \mu, f)$, \textbf{b} $\chi(q, \mu, f)$ and \textbf{c} $U(q, \mu, f)$ with $q = q_c^{MF}(\mu, f)$ versus system size $L$ for $\mu = 0.5$. Lines are linear regressions to the data, and their slopes equal the respective critical exponents in the mean-field limit.}
	\label{expo_mean_field} 
\end{center}
\end{figure*}

Figure \ref{diagram_meanfield} shows the mean-field phase diagram in the $q \times f$ parameter space, revealing the boundary between ordered and disordered phases as a function of $f$ and $\mu$. The lines stand for the analytical solutions given by  \eqref{qc}, and circles represent the numerical data estimates, conforming to mean-field results, summarized in Table \ref{square_lattice_qcs}.

In Fig. \ref{DensityQc}, we use  \eqref{qc} to generate a density plot and visually represent the dynamics of critical noise across the entire spectrum of cooperative behavior parameters $\mu$ and $f$. This plot suggests a symmetry for $q_{c}^{MF}(\mu, f)$ along the line $\mu = 1 - f$. In fact, $q_{c}^{MF}(\mu, f) = q_{c}^{MF}(1 - f, 1 - \mu)$ for all values of $f$ and $\mu$. We conclude that a society characterized by a fraction $f$ of cooperative agents with noise sensibility $\mu$ exhibits equivalent social robustness as a community comprising a fraction $f = 1 - \mu$ of collaborative individuals possessing noise sensitivity $\mu = 1 - f$.

The density plot derived from the mean-field analysis reveals a distinct trend: the rate of growth of the critical social temperature increases sharply as the parameters approach the limits $f \to 1$ and $\mu \to 0$, culminating in a diverging $q_c^{MF}$ at $(f, \mu) = (1, 0)$. We highlight this limit in the white region of Fig. \ref{mxf}, in which the critical social noise would exceed $1$. As a result, society consistently maintains consensus in this region, regardless of the value of $q \leq 1$. Indeed, for $\mu \leq 1/6$, there is a finite critical fraction $f_{c}^{MF}(\mu)$ of cooperative agents in which the system remains ordered for all values of $q$ when $f \geq f_{c}^{MF}(\mu)$. Thus,

\begin{equation}
  f_{c}(\mu \leq 1/6) = \frac{5}{6(1 - \mu)},
\end{equation}

\noindent obtained by using  \eqref{qc} with $q_{c}^{MF}(\mu, f) = 1$.

Finally, we use finite-size scaling relations to plot in Figure \ref{expo_mean_field} (a) magnetization, $(b)$ magnetic susceptibility, and (c) absolute value of the derivative of the Binder cumulant at the critical temperature $q = q_{c}^{MF}(\mu, f)$ versus the system size for $\mu = 0.5$. The line slopes estimate critical exponents $\beta \approx 1/2$, $\gamma \approx 1$, and $\nu \approx 2$ for all values of the investigated $f$ and $\mu$. These results confirm the mean-field cooperative majority-vote dynamics belong to the mean-field Ising universality class.

\section*{Social entropy production}
\label{socioentro}
Entropy production is a manifestation of irreversibility dynamics. The cooperative majority-vote model generates entropy, even in the stationary regime; in contrast, reversible models reach thermodynamic equilibrium states without entropy production in the steady state \cite{hawthorne2023, harunari2019, tome2012entropy}. In this context, we consider the Boltzmann-Gibbs entropy equation at time $t$ 

\begin{equation}
    S(t) = \displaystyle
    - \sum_{\sigma }^{} P(\sigma,t) \text{ ln }P(\sigma,t).
     \label{St}
\end{equation}

\noindent
Combining  \eqref{St} with the master equation of  \eqref{P}, we can express the time derivative of entropy as

\begin{equation}
\begin{aligned}
    \displaystyle \frac{d}{dt}S(t) = \displaystyle
 &   \frac{1}{2}\sum_{\sigma}^{} \sum_{i} \left [w_{i}(\sigma^i) P(\sigma^i,t) -  w_{i}(\sigma) P(\sigma,t)  \right] \\
& \times \text{ ln }\frac{P(\sigma^i ,t)}{ P(\sigma,t)},
     \label{dSt}
     \end{aligned}
\end{equation}

\noindent We frame the rate of change of the entropy $S$ of a system as two main components: entropy production rate $\Pi$ and entropy flux $\Phi$ from system to environment. Thus, we write

\begin{equation}
    \displaystyle \frac{d}{dt}S(t) = \displaystyle \Pi  -\Phi .
     \label{dStzidane}
\end{equation}

\noindent
Therefore, comparing equations \eqref{dSt} and \eqref{dStzidane}

\begin{equation}
     \begin{aligned}
\displaystyle \Pi  = \displaystyle
&    \frac{1}{2}\sum_{\sigma }^{} \sum_{i}  \left [ w_{i}(\sigma^i ) P(\sigma^i ,t) -  w_{i}(\sigma ) P(\sigma,t)  \right ] \\
&\times \text{ ln } \frac{ w_{i}(\sigma^i ) P(\sigma^i ,t) }{w_{i}(\sigma) P(\sigma,t) } ,
     \end{aligned}
     \label{pi}
\end{equation}

\noindent
and 

\begin{figure*}[htpb]
\begin{center}
	\includegraphics[width=1.0\linewidth]{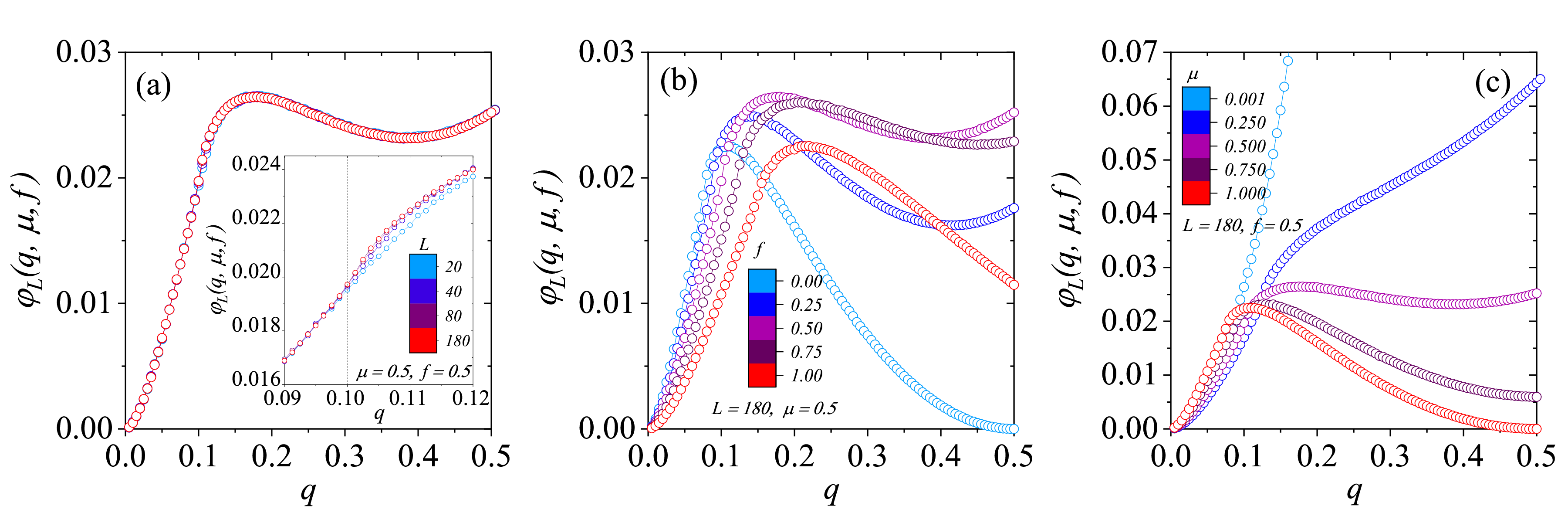} \quad
\caption{\textbf{Stationary social entropy flux production for collaborative majority-vote opinion dynamics versus noise parameter.} We plot $\varphi_{L}(q, \mu, f)$ vs $q$ for several values of \textbf{a} system size $L$, \textbf{b} cooperative fraction $f$ and \textbf{c} noise attenuation $\mu$ on square lattices. On \textbf{a} $\mu = 0.5$ and $f = 0.5$, while in \textbf{b} $\mu = 0.5$ and \textbf{c} $f = 0.5$ with $L = 180$. The lines are guides to the eyes.}
	\label{SoverviewSquareLattice}
\end{center}
\end{figure*}

\begin{equation}
     \begin{aligned}
\displaystyle \Phi  = \displaystyle
&    \frac{1}{2}\sum_{\sigma }^{} \sum_{i}  \left [ w_{i}(\sigma^i ) P(\sigma^i ,t) -  w_{i}(\sigma ) P(\sigma,t)  \right ] \\
&\times \text{ ln } \frac{ w_{i}(\sigma^i )}{w_{i}(\sigma)},
     \end{aligned}
     \label{phi2}
\end{equation}

\noindent Note that $\Pi$ is positive definite, but $\Phi$ can assume either sign depending on the direction of the flux. We write

\begin{equation}
    \displaystyle \Phi  = \displaystyle
    \sum_{\sigma}  \sum_{i }^{} w_{i}(\sigma) P(\sigma,t) \text{ ln } \frac{ w_{i}(\sigma )  }{ w_{i}(\sigma^i)},
     \label{phi}
\end{equation}
that allows numerical estimates \cite{nicolis1977self, lebowitz1999gallavotti, maes1999fluctuation,maes2000definition,maes2003time, lecomte2005energy}.

\subsection*{Flux on square lattices}
\begin{figure*}[ht!]
\begin{center}
	\includegraphics[width=0.7\linewidth]{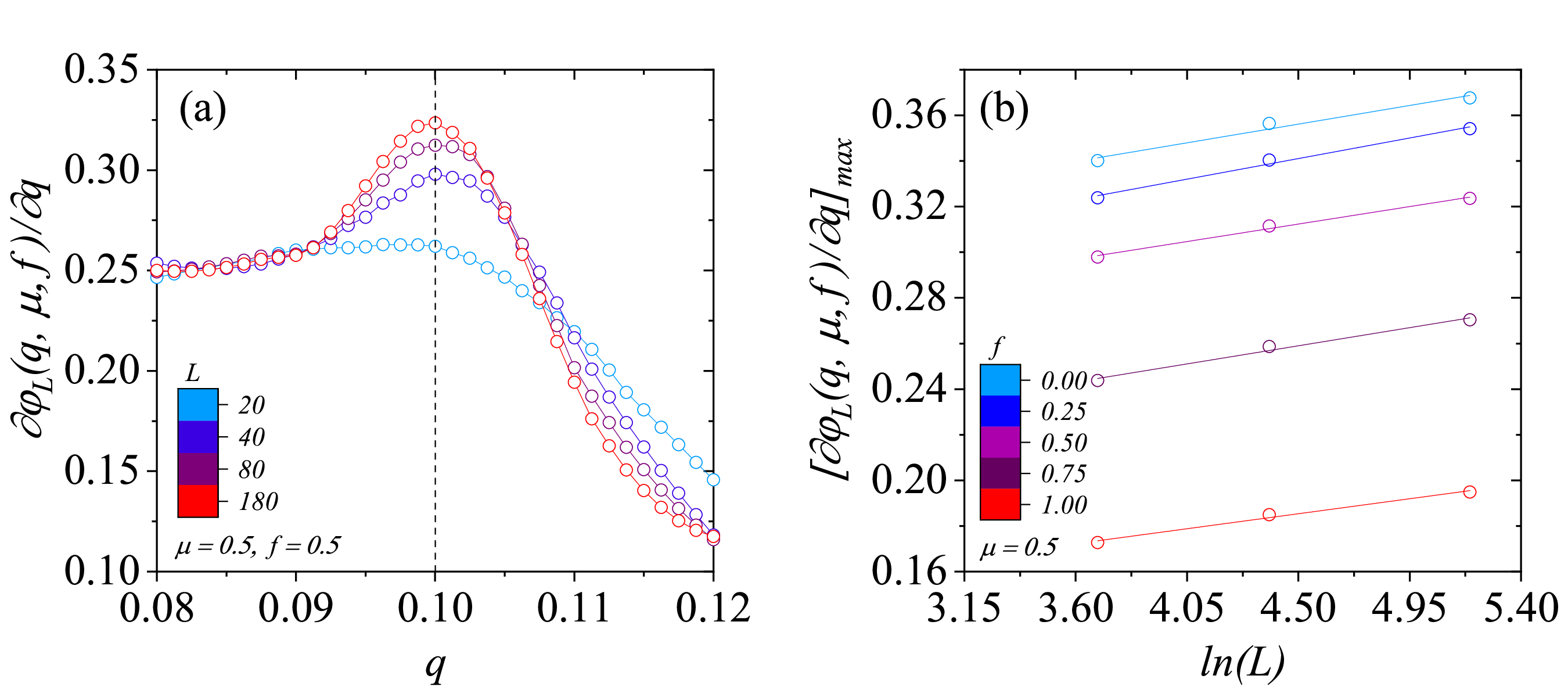} \quad
\caption{\textbf{Entropy flux Size dependence.} \textbf{a} Derivative of entropy flux versus $q$ for $\mu = 0.5$ and $f = 0.5$. \textbf{b} shows the maximum value of the entropy derivative at the critical point as a function of the natural logarithmic of the system size $L$. From top to bottom, line slopes are $\eta = 0.018(3), 0.020(2), 0.017(1), 0.018(2)$ and $0.015(2)$.}
	\label{PhiDerivative}
\end{center}
\end{figure*}

The flux of entropy as a configurational average over the probability distribution in the stationary state from  \eqref{phi} is 

\begin{equation}
    \displaystyle \Phi  = \displaystyle
    \sum_{i} \left \langle  w_{i}(\sigma)  \text{ ln } \frac{ w_{i}(\sigma )  }{ w_{i}(\sigma^i)} \right \rangle.
    \label{phimean}
\end{equation}
The social entropy $S$ remains constant in this state, therefore $\Pi = \Phi$. Hence, we calculate the stationary social entropy production by employing Monte Carlo simulations using \eqref{phimean}.

In Fig. \ref{SoverviewSquareLattice}, we plot numerical results of entropy production $\varphi_{L}(q, \mu, f)$ in the stationary regime for several values of (a) system size $L$, (b) collaborative fraction $f$ and (c) noise attenuation $\mu$ versus $q$. We observe in (a) that entropy flux has a weak sensibility with population size but a strong dependence on the fraction of collaborative agents and noise attenuation. Curves for $f=0.0$ and $\mu=1.0$ in Figures \ref{SoverviewSquareLattice}(b) and (c), respectively, display the flux of entropy of the isotropic MVM \cite{de1992isotropic, de1993isotropic}, where a maximum flux occurs after the critical noise $q_{c}(\mu, f = 0.0) = 0.075$ and tends to zero for $q \to 0$ or $q \to 1/2$ in the isotropic case.

We highlight Fig. \ref{SoverviewSquareLattice}(b) also displays entropy flux that follows the isotropic system under the linear transformation $q \to q/\mu$, when $f=1.00$ and $\mu = 0.5$. This flux vanishes for $q \to 0$ and $q \to 1/2\mu$. For $0 < f <1$, after the maximum, instead of approaching zero, the entropy flux increases, supported by a discrepancy between the behavior of the cooperative and regular individuals. Indeed, (c) reveals this phenomenon intensifies as $\mu$ becomes smaller since the social temperature disparity among agents increases. We remark that, in general, a combination of cooperative and non-cooperative individuals increases the social entropy production of the society. Nonetheless, for small values of the social noise, the entropy generation is maximized when there is no cooperative behavior, $f=0.0$.

As a general pattern, the critical temperature does not coincide with the maximum of $\varphi_{L}(q, \mu, f)$. In fact, the critical noise is the point of inflection for $\varphi$ that occurs before the maximum point. We display this behavior in the inset of Fig. \ref{SoverviewSquareLattice}(a), in which for $\mu = 0.5$ and $f = 0.5$, we obtain $q_c = 0.1011(1)$ (see Table \ref{square_lattice_qcs}). Therefore, in analogy with the entropy of equilibrium Ising model, the entropy flux exhibits a finite singularity at the critical point as

\begin{equation}
    \displaystyle
     \varphi_{L}(q, \mu, f) = \varphi_{L}[q_{c}, \mu, f] + A_{\pm}\left | q - q_{c} \right| ^{(1 - \alpha)},
      \label{singularity}
\end{equation}
where $A_{\pm}$ are amplitudes of regimes above and under the critical point $q_c = q_{c}(\mu, f)$ \cite{hawthorne2023, harunari2019}. Hence, instead of a maximum in entropy flux, the second-order phase transition maximizes the derivative of entropy flux with respect to $q$, as we can observe in Fig. \ref{PhiDerivative}(a) for $q_{c}(0.5, 0.5) = 0.1011(1)$. Indeed, from  \eqref{singularity}, we obtain

\begin{equation}
    \displaystyle
     \frac{\partial \varphi_{L}(q, \mu, f)}{\partial q} \sim \left | q - q_{c} \right|^{-\alpha},
      \label{singularityDerivative1}
\end{equation}
where $\alpha$ corresponds to the same exponent associated with the specific heat of the Ising model. On square lattices, $\alpha = 0$, generating a singularity of the logarithm type. Hence, in analogy to the Ising model, we write
\begin{equation}
    \displaystyle
     \frac{\partial \varphi_{L}(q, \mu, f)}{\partial q} \sim \text{ln}\left | q - q_{c} \right|.
      \label{singularityDerivative2}
\end{equation}
To verify our conjecture, we use the Savitzky-Golay Smooth algorithm with cubic polynomials to numerically estimate $\partial \varphi_{L}(q, \mu, f)/\partial q$ for several sizes $L$ in Fig. \ref{PhiDerivative}(a). By finite-size scaling theory on  \eqref{singularityDerivative2}, the maximum value of the partial derivative of entropy flux must diverge at the critical point as

\begin{equation}
   \displaystyle
     \left [ \frac{\partial \varphi_{L}(q, \mu, f)}{\partial q} \right ]_{max} \sim \text{ln } L^{\eta},
      \label{singularityDerivative3}
\end{equation}
with $\eta = (1-\zeta)/\nu$ and $\nu$ is the critical exponent associated to correlation length. Indeed, our results support $\nu = 1.0$, leading to $\zeta \approx 1.0$. Figure \ref{PhiDerivative}(b) confirms our assumption for $\mu = 0.5$ and several cooperative fraction values. We observe the same behavior for other values of $\mu$ and $f$.

\subsection*{Mean-field approach}
The mean-field theory allows us to develop an analytical expression for entropy flux in the stationary regime. From  \eqref{w}, we write

\begin{equation*}
    \displaystyle 
     \text{ ln} \frac{w_{i}(\sigma)}{w_{i}(\sigma^i)} = \text{ ln}\left [\frac{ 1 - \sigma _{i} S \left ( \sum_{\delta} \sigma_{i+\delta  } \right ) + 2\mu_i q \sigma _{i} S \left ( \sum_{\delta} \sigma_{i+\delta  } \right ) }{ 1 + \sigma _{i} S \left ( \sum_{\delta} \sigma_{i+\delta  } \right ) - 2\mu_iq \sigma _{i} S \left ( \sum_{\delta} \sigma_{i+\delta  } \right ) }\right ].
\end{equation*}
Next, we note that the product $\sigma _{i} S \left ( \sum_{\delta} \sigma_{i+\delta  } \right )$ may assume only one of three possible values: $-1, 0$ and $1$. Therefore,
\begin{equation*}
    \displaystyle 
     \text{ ln} \frac{w_{i}(\sigma)}{w_{i}(\sigma^i)} =  \left\{\begin{matrix}
\text{ln}  \left ( \frac{\mu_i q}{1-\mu_{i}q} \right )\times 1, & \text{if } \sigma _{i} S \left ( \sum_{\delta} \sigma_{i+\delta  } \right )= 1,  \\
 \text{ln}  \left ( \frac{\mu_i q}{1-\mu_{i}q} \right )\times 0, & \text{if } \sigma _{i} S \left ( \sum_{\delta} \sigma_{i+\delta  } \right )= 0,  \\
 \text{ln}  \left ( \frac{\mu_i q}{1-\mu_{i}q} \right )\times -1, & \text{if } \sigma _{i} S \left ( \sum_{\delta} \sigma_{i+\delta  } \right )= -1.  \\
\end{matrix}\right.
\end{equation*}
Thus, we obtain
\begin{equation}
     \text{ln} \frac{ w_{i}(\sigma)}{w_{i}(\sigma^i)} = \text{ln}  \left [ \frac{\mu_i q}{1-\mu_{i}q} \right ]\sigma_{i} S\left (\sum_{\delta} \sigma_{i + \delta } \right ).
     \label{lnwwi}
\end{equation}
Combining the equations \eqref{phimean} and \eqref{lnwwi}
\begin{equation*}
    \displaystyle 
    \displaystyle \Phi = \sum_{j =1}^{Nf} \left \langle \text{ln} \left [ \frac{\mu q}{1 -\mu q} \right ]  \sigma_{j} S\left (\sum_{\delta} \sigma_{j + \delta} \right )w_{j}(\sigma) \right \rangle +
\end{equation*}

\begin{equation}
    \displaystyle 
     + \sum_{k = Nf+1 }^{N} \left \langle \text{ln}\left [\frac{q}{1-q} \right]\sigma_{k}S\left ( \sum_{\delta}  \sigma_{k + \delta} \right )w_{k}(\sigma)  \right \rangle.
     \label{phimodel}
\end{equation}
Furthermore, in the stationary state, we obtain 
\begin{equation}
     \left< \left [ S \left (\sum_{\delta} \sigma_{j + \delta } \right ) \right ]^2 \right> = \frac{1}{8}\left ( 5 +6m^2 -3m^4 \right ).
     \label{S2}
\end{equation}
We divide  \eqref{phimodel} by the total number of individuals $N$ and combine it with  \eqref{w} and \eqref{S2} to derive an expression for entropy flux per site:

\begin{equation}
\begin{aligned}
    \varphi  \equiv  \Phi/N = f \text{ ln}\left (\frac{\mu q}{1 -\mu q} \right ) \times \\\left[\frac{1}{4} (3m^2-m^4) - \frac{\Theta_{\mu}}{16}(5 + 6m^2 -3m^4)\right] + \\
    + \displaystyle
      (1-f) \text{ ln}\left ( \frac{ q}{1-q} \right ) \times \\
      \left[\frac{1}{4} (3m^2-m^4) - \frac{\Theta}{16}(5 + 6m^2 -3m^4)\right].
      \end{aligned}
     \label{phiexpression}
\end{equation}

\noindent
We set $m = 0$ and get the disordered solution of the entropy flux, valid for $q > q_{c}^{MF}(\mu,f)$:
\begin{equation}
\begin{aligned}
    \displaystyle
     \varphi  = \frac{5}{16} f \Theta_{\mu}   \text{ ln} \left ( \frac{  1 -\mu q   }{\mu q} \right )  + \\
     + \frac{5}{16}(1-f) \Theta  \text{ ln}\left ( \frac{1- q  }{q} \right ).
     \end{aligned}
     \label{phiparamagnetic}
\end{equation}

\noindent
On the ordered state, we have that magnetization behaves accordingly to  \eqref{p8}, which is valid for $q < q_{c}^{MF}(\mu,f)$. Combining  \eqref{p8} and \eqref{phiexpression}, we obtain the entropy flux expression in the ferromagnetic state:

\begin{equation*}
    \displaystyle
     \varphi = \frac{f}{(1 - 2\bar q)^2} \text{ ln}  \left ( \frac{ 1-\mu q}{\mu q} \right ) \left \{\bar q \left [3 - \Theta_\mu(2 + \bar q) \right] - \mu q \right \} + 
\end{equation*}

\begin{equation}
    \displaystyle
     + \frac{1-f}{(1 - 2\bar q)^2} \text{ ln}  \left ( \frac{ 1-q}{q} \right ) \left \{\bar q\left [3 - \Theta(2 + \bar q) \right] - q \right \},
      \label{phiferromagnetic}
\end{equation}

\noindent with $\bar{q} = \bar{\mu}q = q[(1-f( 1-\mu)]$. For the particular case $f = 0.0$, we combine  \eqref{phiparamagnetic} and \eqref{phiferromagnetic} to obtain an expression for entropy production of the isotropic majority-vote model:
\begin{equation*}
    \varphi(q) = \left ( \frac{q}{1-2q} \right )^2 (3+2q)\text{ ln} \left ( \frac{1-q}{q} \right )H(q_{c} - q) + 
\end{equation*}
\begin{equation}
    +\frac{5}{16}(1-2q)\text{ ln} \left ( \frac{1-q}{q} \right )H(q - q_{c}),
    \label{phiStandardMVM}
\end{equation}
where $H(t)$ is the Heaviside function and $q_{c}^{MF}(\mu, f=0.0) = 1/16$ is the isotropic mean-field MVM critical noise. We further investigate numerical simulations in the mean-field formulation to demonstrate this result. 

\begin{figure}[htpb]
\begin{center}
	\includegraphics[width=1.0\linewidth]{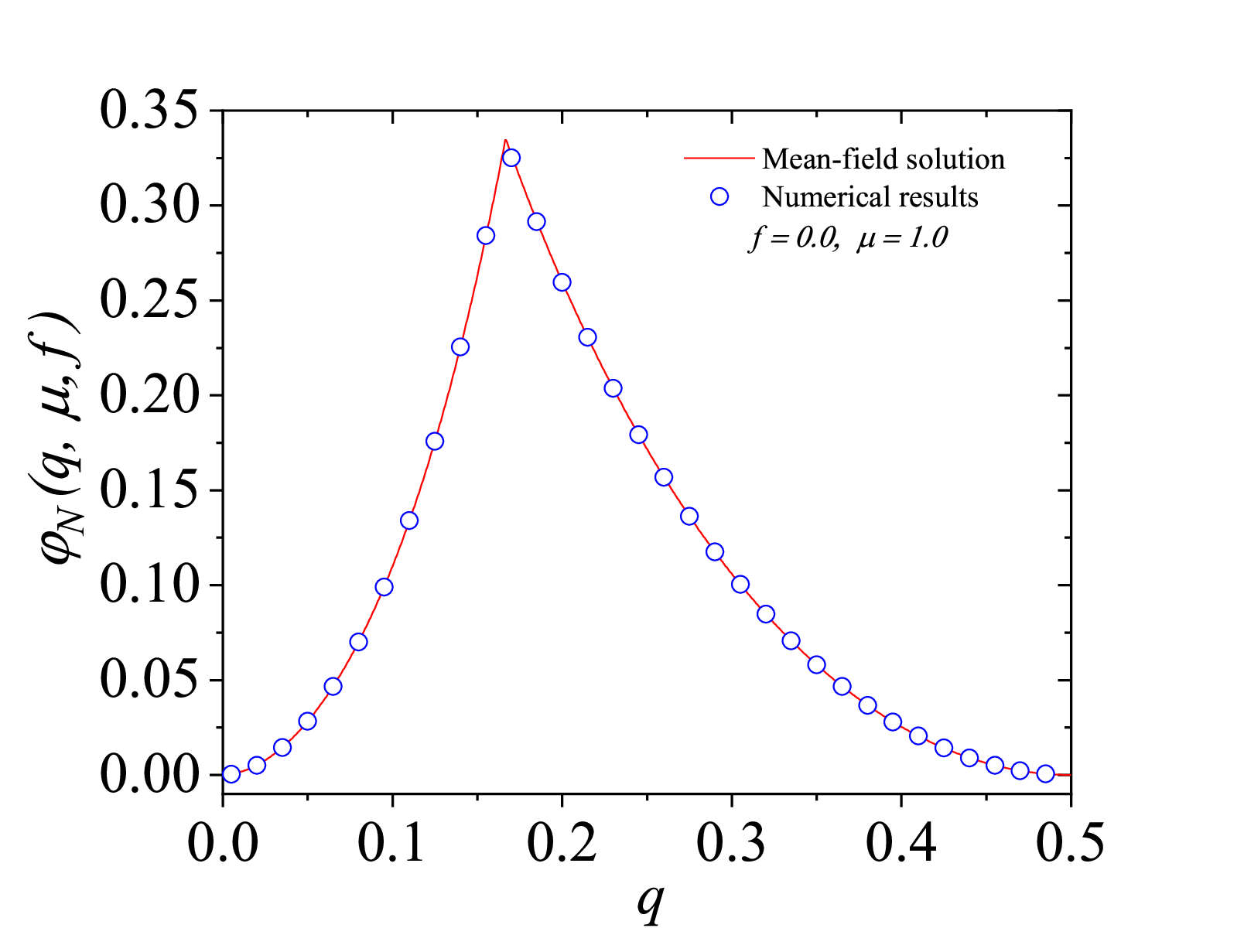} \quad
\caption{\textbf{Entropy production for mean-field isotropic majority-vote model.} The red line denotes the results from \eqref{phiStandardMVM} and blue circles are numerical estimations and $N = 32400$.}
	\label{Tania}
\end{center}
\end{figure}

\begin{figure*}[htpb]
\begin{center}
 	\includegraphics[width=\linewidth]{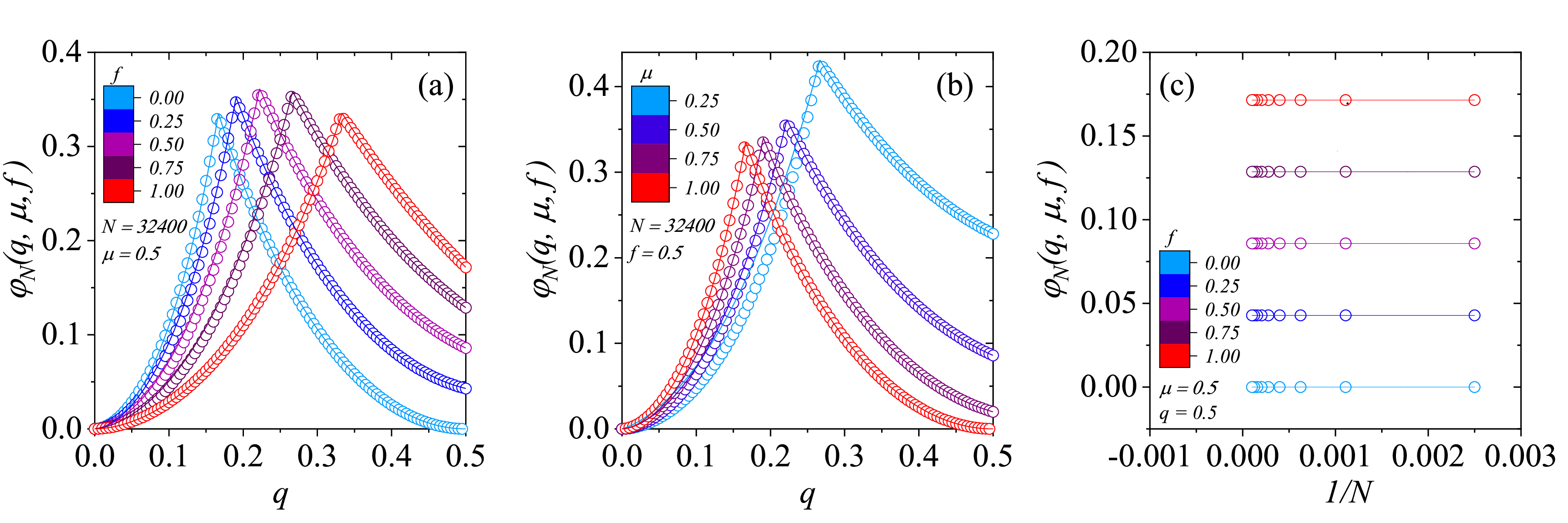} \quad
\caption{\textbf{Mean-field stationary social entropy flux production as a function of the social temperature.} \textbf{a} Flux production dependence on cooperative fraction $f$ for $\mu = 0.5$, \textbf{b} and several values of noise sensibility $\mu$ with $f = 0.5$. Open circles are numerical data from mean-field Monte Carlo simulations with $N = 32400$, and lines represent analytical results of \eqref{phiparamagnetic} and \eqref{phiferromagnetic}. \textbf{c} illustrates the finite behavior of entropy flux production for $q = 0.5$ which is independent of system size.}
	\label{meanfield_phi}
\end{center}
\end{figure*}

Figure \ref{Tania} reveals mean-field entropy flux $\varphi(q, \mu, f)$ versus $q$, with $N = 32400$, for the isotropic mean-field MVM simulation ($f = 0$ and $\mu = 1.0$). The red line represents results from  \eqref{phiStandardMVM}, where blue circles stand for numerical data in the mean-field formulation. Note that $\varphi$ exhibits a singularity in mean-field critical noise $q_{c}^{MF} = 1/6$ and vanishes for $q \to 0$ and $q \to 1/2$. 

We extend our investigation for mean-field stationary social entropy flux production $\varphi(q, \mu, f)$ versus noise $q$ for several values of the cooperative fraction $f$ and noise sensibility $\mu$. In Fig. \ref{meanfield_phi}(a), we set $\mu = 0.5$ and $f = 0.00, 0.25, 0.50, 0.75$ and $1.00$, while in the Fig. \ref{meanfield_phi}(b), $f = 0.5$ and $\mu$ assume $0.25, 0.50, 0.75$ and $1.00$. The open circles are mean-field numerical data for $N = 32400$ individuals, and the lines represent the analytical results given by  \eqref{phiparamagnetic} and \eqref{phiferromagnetic}. There are slight deviations between mean-field solutions and Monte Carlo data in the ferromagnetic phase for entropy flux due to the finite nature of simulated systems, amplified as $\mu \to 0$.

Figure \ref{meanfield_phi}(c) shows that $\varphi$ does not approach zero when $q = 0.5$ in the mean-field limit for $0 < f < 1$ but remains finite independently of system size $N$. However, for any isotropic case ($f = 0$ or $\mu = 1.0$), $\varphi$ tends to zero for $q = 0.5$, as expected. We also highlight that in the mean-field framework, the critical temperature $q_{c}^{MF}(\mu, f)$ corresponds to the maximum entropy flux point.
\section*{Discussion}
\label{conc}

This work explores the impacts of collaborative behavior on majority-vote opinion dynamics and its social entropy production. We randomly select a fraction $f$ of individuals of the society to represent cooperative agents, while the complementary fraction $1-f$ are regular voters. We introduce a social noise $q$ such that with probability $(1-q)$, individuals agree with each other regarding a social issue subject, such as a political, professional, or economic matter. The cooperative agents retain a social temperature sensibility $0 < \mu < 1$, experiencing an effective social noise of $\mu q$, favoring social validation-based decisions. For $\mu = 0$ and $\mu = 1$, we recover bimodal \cite{vilela2012majority, vilela2017small} and isotropic majority-vote model \cite{de1992isotropic, de1993isotropic}, respectively. 

We employ Monte Carlo simulations and find that the system undergoes a second-order dissensus-consensus phase transition with the same universality class of 2D equilibrium Ising model for critical noise values $q = q_c(\mu, f)$. The critical exponents are not affected by the presence of collaborative agents, following Grinstein’s criterion which states that nonequilibrium stochastic spin-like systems with up-down symmetry in regular lattices fall into the same universality class of the equilibrium Ising model \cite{grinstein1985statistical,baxter1982inversion}.  

For heterogeneous societies ($0 < f < 1$), there is a contrast between the effects of social temperature among regular and cooperative individuals, and $q_c(\mu, f)$ is a monotonically decreasing (increasing) function of noise attenuation $\mu$ (cooperative fraction $f$). Indeed, increasing the cooperative fraction $f$ promotes the formation of a giant cluster of agreeing individuals that suppresses the phase transition. The collaborative behavior phenomena enhance the social robustness of society to opinion polarization. We highlight that if all individuals are cooperative ($f = 1.0$), the system behaves as if all individuals were regular ($f = 0.0$) under the linear transformation $q \longrightarrow q/\mu$.  

Gibbs entropy and the master equation yield an expression that enables us to compute social entropy production in the stationary regime for square lattices. We observe that the entropy production of the isotropic majority-vote model has a maximum that occurs after the critical noise $q_{c}(\mu, f = 0.0) = 0.075$ and vanishes for $q \to 0$ or $q \to 1/2$. However, for cooperative societies, the entropy production increases after the local maximum and is non-zero for $q = 1/2$ due to the social temperature disparity between the collaborative and regular agents. Furthermore, combining cooperative and non-cooperative individuals yields higher social entropy production. Yet, for small social noise values, maximum entropy generation occurs in the absence of cooperative behavior, with $f=0.0$. 

Further generalizations of heterogeneous majority-vote opinion dynamics may consider complex network framework and the presence of non-compliance agents, in other words, $\mu > 1$. Exploring the influence of these dissenting agents within opinion dynamics can provide a deeper understanding of societal and (non)cooperative behaviors.

\acknow{We thank M. J. de Oliveira for a critical reading of the manuscript. The authors acknowledge financial support from Brazilian and Chinese institutions and funding agents UPE, FACEPE (APQ-0565-1.05/14, APQ-0707­-1.05/14), CAPES, CNPq (306068/2021-4), National Natural Science Foundation of China (72071006, 61603011, 62073007). The Boston University Center for Polymer Studies is supported by NSF Grants PHY-1505000, CMMI-1125290, and CHE-1213217, by DTRA Grant HDTRA1-14-1-0017, and by DOE Contract DE-AC07-05Id14517.}

\showacknow{} 

\bibliographystyle{unsrtnat}
\bibliography{References}

\end{document}